\documentclass[twocolumn,amsfonts,aps,prc,nofootinbib,floatfix]{revtex4}

\usepackage{amsmath}
\usepackage{graphicx}

\def\HoTCoffeeh{{\fontfamily{cmtt}\selectfont HoTCoffeeh} }

\def\l{\left}
\def\r{\right}

\def\ev{\mathrm{ev}}

\begin{document}

\title{The Shape of the Correlation Function}

\author{Jakub Cimerman}
\affiliation{\v{C}esk\'e vysok\'e u\v{c}en\'i technick\'e v Praze, FJFI, B\v{r}ehov\'a 7, 11519 Prague, Czechia}

\author{Christopher Plumberg}
\affiliation{Theoretical Particle Physics, Department of Astronomy and Theoretical Physics, Lund University, S\"olvegatan 14A, SE-223 62 Lund, Sweden}

\author{Boris Tom\'a\v{s}ik}
\affiliation{Univerzita Mateja Bela, FPV, Tajovsk\'eho 40, 97401 Bansk\'a Bystrica, Slovakia\\ and
\v{C}esk\'e vysok\'e u\v{c}en\'i technick\'e v Praze, FJFI, B\v{r}ehov\'a 7, 11519 Prague, Czechia}

\begin{abstract}
The two-particle correlation function employed in Hanbury-Brown Twiss interferometry and femtoscopy is traditionally parameterized by a Gaussian form.  Other forms, however, have also been used, including the somewhat more general L\'evy form.  Here we consider a variety of effects present in realistic femtoscopic studies which may modify the shape of the correlation function and thereby influence the physical interpretation of a given parameterization.
\end{abstract}

\date{\today}

\maketitle

\section{Introduction}

The observables of Hanbury Brown-Twiss (HBT) interferometry and femtoscopy%
\footnote{While femtoscopy can be performed for any identified pairs of particles, we here reserve the acronym ``HBT" for the use of identical hadrons, which can also be performed on proton pairs and kaon pairs.} 
have become a standard part of the experimental arsenal for measuring and probing the space-time evolution of heavy-ion collisions.  By construction, these observables use the presence of quantum statistical correlations between identical particles in momentum-space to infer aspects of the spatio-temporal structure of the freeze-out surface \cite{Heinz:1999rw,Lisa:2005dd}.  The nature of this inference depends largely upon the way in which the momentum-space correlations are parameterized in their relation to the space-time structure of the emitting source.

For an emitting source whose spatial density falls off at large distances faster than $r^{-2}$, one can show that the leading behavior of this parameterization should be Gaussian in form \cite{Heinz:1996bs}.  Nevertheless, several alternatives to the traditional leading Gaussian parameterization have been considered in the literature, some of which fit the data quite well \cite{Adare:2017vig}.  In order to understand why this may be the case and further assess the meaning of a given parametrization, it is therefore crucial to assess the ways in which it can be influenced at the quantitative level by various physical effects present in realistic measurements of heavy-ion collisions.

One way to assess the influence of such effects, in practice, is by means of detailed quantitative modeling.  This is the approach we adopt in this paper.  Our goal will be to assess the importance of various physical effects which may influence the source characteristcs extracted from a particular parameterization known as the L\'evy distribution.  In particular, we will be interested in determining whether a Gaussian source of particles 
can `mimic' a more general L\'evy parametrization of the correlation function under appropriate circumstances.

The task to understand the physics behind the L\'evy parametrisation is very important in view of the current search for the critical point in the QCD phase diagram. It has been suggested that the produced matter in the vicinity of the critical point would result in a L\'evy stable correlation function with L\'evy index (defined below) $\alpha = 1/2$ \cite{Csorgo:2009gb}. Here, we investigate how also other non-critical phenomena may influence the value of $\alpha$.

To answer this question, we begin in Sec.~\ref{Formalism} by discussing the general formalism for parametrizing and extracting physics from the two-particle correlation functions which measure quantum statistical correlations.  We first consider both Gaussian and L\'evy parametrizations of the correlation function in general, and then we proceed to consider the ways in which the former can be distorted to look more like the latter in realistic situations.  The sources of distortion which we focus on in this work include ensemble averaging, one-dimensional representations of the correlation function, averaging over finite bins in the pair momentum $\vec{K}$, and resonance decay effects.

A quantitative analysis of these various sources requires detailed model studies, so in Sec.~\ref{ModelStudies} we describe two different models that we use for simulating heavy-ion collisions and computing the associated correlation functions.  These models are based respectively on the blast-wave model (Sec.~\ref{BlastWaveModel}) and on hydrodynamics (Sec.~\ref{Hydrodynamics}), both of which have proven to be indispensable tools in the analysis of heavy-ion collisions.  We will find in Sec.~\ref{Results} that the primary qualitative conclusions we wish to draw from these results do not depend significantly on which model is used.  We conclude in Sec.~\ref{Conclusions} by reiterating our most important results.

\section{Formalism}
\label{Formalism}

HBT is a powerful technique for probing geometric properties of an emitting source through the analysis of momentum-space correlations \cite{Heinz:1999rw,Lisa:2005dd}.  In addition to being used in probing heavy-ion collisions, it was originally applied in the context of astronomy \cite{HanburyBrown:1954amm,Brown:1956zza,HanburyBrown:1956bqd}.  In this section, we will present the most important elements of standard HBT analyses and outline the way in which space-time information about the emitting source may be extracted by analyzing the two-particle correlation function.  This will then lead us to discuss some of the ways in which the usual treatment can be complicated by various aspects of a realistic analysis.

\subsection{Basics of HBT femtoscopy}

The starting point of HBT is the two-particle correlation function which probes the momentum-space structure of correlations between pairs of identical particles produced in heavy-ion collisions.  In this study, we focus exclusively on correlation functions constructed from charged pion pairs. The correlation function is defined by \cite{Wiedemann:1999qn,Lisa:2008gf}
\begin{equation}
	C( \vec{p}_1, \vec{p}_2 )
		= \frac{E_1 E_2 d^6N/\l( d^3p_1 d^3p_2 \r)}{\l( E_1 d^3N/d^3p_1 \r) \l( E_2 d^3N/d^3p_2 \r)}
	\label{correl_func_def}
\end{equation}
In the absence of physical correlations, $C\l( \vec{p}_1, \vec{p}_2 \r) = 1$.

Theoretically, the correlation function can also be approximately formulated in terms of the Wigner distribution (also known as emission function) governing particle emission at  freeze-out:
\begin{eqnarray}
	C\l( \vec{q}, \vec{K} \r)
		&=& 1 + \frac{\l| \int d^4x\, e^{i q \cdot x} S(x,K) \r|^2}{\l( \int d^4x\, S(x,K+\frac{q}{2}) \r) \l( \int d^4x\, S(x,K-\frac{q}{2}) \r)} \nonumber\\
		&& \label{correl_func_FT_S_no_smooth}\\
		& \approx & 1 + \frac{ \l| \int d^4x\, e^{i q \cdot x} S(x,K) \r|^2 }{ \l( \int d^4x\, S(x,K) \r)^2}\,,
	\label{correl_func_FT_S}
\end{eqnarray}
where $q^\mu = p_1^\mu - p_2^\mu$ is the relative momentum, $K^\mu = \l(p_1^\mu + p_2^\mu \r)/2$ is the pair momentum, and the relation
\begin{equation}
	q \cdot K = 0 \label{onshell_condition}
\end{equation}
constrains individually measured pions to be on shell \cite{Heinz:2004qz}.

The validity of the second step in \eqref{correl_func_FT_S} relies on the so-called ``smoothness approximation," which is justified for sufficiently large systems such as heavy-ion collisions \cite{Pratt:1997pw}.  For smaller collision systems, which we do not consider in this work, one needs to revert back to \eqref{correl_func_FT_S_no_smooth}.

Space-time information regarding the emitting source can then be obtained by parametrizing the Bose-Einstein enhancement term in an appropriate fashion.  One conventional and well-motivated choice is the Gaussian form \cite{Heinz:1996bs}\footnote{We assume that the data have already been corrected for the effects of Coulomb interactions \cite{Wiedemann:1998ng}.}
\begin{equation}
C_{\mathrm{G}}( \vec{q}, \vec{K} )
	= 1 + \lambda( \vec{K} )
			\exp\l[ - \sum_{i,j=o,s,l} R^2_{ij}( \vec{K} ) q_i q_j \r]\,    .
	\label{correl_func_Gaussian_form}
\end{equation}
The indices $o$, $s$, $l$ refer to the \emph{out-side-longitudinal} reference frame. The longitudinal axis coincides with the beam direction, while the outward axis is parallel to the transverse component of the pair momentum $K_T$. The sideward direction is then perpendicular to these two. 

Parametrized in this way, the correlation function contains seven undetermined parameters: the intercept parameter $\lambda( \vec{K} )$ and the HBT radii $R^2_{ij}( \vec{K} )$.  In general, $\lambda$ ranges between 0 and 1, and quantifies the magnitude of the Bose-Einstein enhancement when $\vec{q} = 0$ \cite{Wiedemann:1999qn}.  The $R^2_{ij}$, on the other hand, reflect the space-time information implicit in the correlation function \eqref{correl_func_def}.  In particular, the $R^2_{ij}( \vec{K} )$ can be understood in terms of ``homogeneity lengths" \cite{Akkelin:1995gh} characterizing the regions of the freeze-out surface which predominately emit particles with average momentum $\vec{K}$.

The correlation function can in principle be measured fully differentially in $\vec{q}$ and $\vec{K}$, and several studies carried out along these lines have yielded valuable insights into important aspects of heavy-ion collisions \cite{Khachatryan:2010un,Aamodt:2010jj,Aamodt:2011kd,Abelev:2014pja}.  Such analyses require a relatively large number of pion pairs in order to make an accurate measurement of \eqref{correl_func_def}.  For this reason, it is conventional to modify the basic analysis described above in a number of ways which, while improving the statistical precision of the measurement, implicitly involve different lengths scales in the problem and thus give reasons for a non-Gaussian shape of the correlation function.
It is, moreover, well known that the Gaussian form is further altered, even in a fully differential measurement of \eqref{correl_func_def}, by the contribution to the final pion yield coming from resonance decays.  Hereafter, we will often refer to these effects collectively as `non-Gaussian effects' or `non-Gaussian sources'.  Clearly, the presence of such non-Gaussian sources can lead to significant deviations from the traditional Gaussian form \eqref{correl_func_Gaussian_form}.

It is therefore important to take note of alternatives to the standard Gaussian parameterization which have been proposed in the literature.  Examples include exponential forms \cite{Khachatryan:2010un,Aamodt:2010jj}, various types of power laws \cite{Eggers:2005qz}, Edgeworth expansions \cite{Csorgo:1999wx,Abelev:2014pja}, and even mixtures of exponentials and Gaussians \cite{Aamodt:2011kd}.  Another popular, somewhat more general form is the L\'evy (or L\'evy-stable) distribution, which has been explored in connection with sensitivity to enhanced fluctuations near a QCD critical point \cite{Csorgo:2009gb,Adare:2017vig}.
The Gaussian and exponential parameterizations are special cases of the L\'evy distribution, meaning that the latter provides one way of quantifying deviations in the measured correlation function from exact Gaussian or exponential scaling.

More precisely, the L\'evy distribution is defined by the following generalization of the Gaussian and exponential parameterizations \cite{Csorgo:2003uv}:
\begin{equation}
C_{\mathrm{L}}( \vec{q} )
	= 1 + \lambda
			\exp\l[ - \l| \sum_{i,j=o,s,l} R^{\prime 2}_{ij} q_i q_j\r|^{\alpha/2} \r]\,,
	\label{correl_func_Levy_form}
\end{equation}
and we have suppressed the explicit dependence on $\vec{K}$.  Here, $\alpha$ is typically referred to as the L\'evy exponent, and the ${R'}^2_{ij}$ parameters are analogous to the $R^2_{ij}$ appearing in \eqref{correl_func_Gaussian_form}.  If the cross terms in \eqref{correl_func_Levy_form} are neglected, we obtain
\begin{equation}
C_{\mathrm{L}}( \vec{q} )
	= 1 + \lambda
			\exp\l[ - \l( R^{\prime 2}_o q_o^2 + R^{\prime 2}_s q_s^2 + R^{\prime 2}_l q_l^2 \r)^{\alpha/2} \r]\,.
	\label{correl_func_Levy_3Ddiag_form}
\end{equation}
Finally, the one-dimensional case can be written
\begin{equation}
C_{\mathrm{L}}( Q )
	= 1 + \lambda
			\exp\l( -\l| R' Q \r|^{\alpha} \r)\,,
	\label{correl_func_Levy_1D_form}
\end{equation}
where $Q$ is a scalar version of the relative momentum $q$ which will be defined below.

The L\'evy exponent $\alpha$ characterizes the shape of the correlation function,
and thus provides one method for testing the `degree of non-Gaussianity' present in the data.  While some of this non-Gaussianity can originate from exotic phenomena like critical fluctuations, it is important to assess the extent to which more mundane effects, such as the contributions from resonance decays mentioned above, can influence the shape of the observed correlation function as well.  It is therefore necessary to understand how these latter effects can distort correlations present in an otherwise Gaussian source, in order to establish a baseline for interpreting $\alpha$ in terms of genuinely non-Gaussian features of the correlation function.  In the remainder of this section, we describe several of these effects in greater detail and the way in which they must be evaluated quantitatively in modeling HBT and femtoscopic analyses.

\subsection{Ensemble averaging}

In order to build up statistics, it is conventional to average the numerator and denominator of \eqref{correl_func_def} separately over a large collection of collision events.  This procedure is often referred to as either `ensemble averaging' or `event averaging'.  We use the two terms interchangeably here.

After averaging, \eqref{correl_func_def} and \eqref{correl_func_FT_S} must be replaced by \cite{Plumberg:2013nga,Plumberg:2015mxa}
\begin{equation}
	C_{\mathrm{ev.avg.}}\l( \vec{p}_1, \vec{p}_2 \r)
		= \frac{\l< E_1 E_2 d^6N/\l( d^3p_1 d^3p_2 \r) \r>_{\ev}}{\l< E_1 d^3N/d^3p_1 \r>_{\ev} \l< E_2 d^3N/d^3p_2 \r>_{\ev}}
	\label{avg_correl_func_def}
\end{equation}
and
\begin{eqnarray}
	C_{\mathrm{ev.avg.}}( \vec{q}, \vec{K} )
		& \approx & 1 + \frac{\l< \l| \int d^4x\, e^{i q \cdot x} S(x,K) \r|^2 \r>_{\ev}}{\l< \l( \int d^4x\, S(x,K) \r)^2 \r>_{\ev}}\,.
	\label{avg_correl_func_FT_S}
\end{eqnarray}

\subsection{$\vec{K}$ averaging}
The size of a bin in the pair momentum $\vec{K}$ clearly cannot be taken to be arbitrarily small for a fixed number of pion pairs.  This implies that correlation functions must be averaged over some pair momentum interval whose size is determined by the density of available pairs.  Since a specific momentum is usually  produced just by a part of the whole fireball---called the homogeneity region---taking an interval of pair momenta implies averaging over different homogeneity regions. Following similar logic as in the previous subsection, the averaged correlation function can be written schematically as
\begin{eqnarray}
	C( \vec{q}, \vec{K} )
		& \approx & 1 +
			\frac{\int_{\mathrm{bin}} d^3K\, \l| \int d^4x\, e^{i q \cdot x} S(x,K) \r|^2 }
			{ \int_{\mathrm{bin}} d^3K\, \l( \int d^4x\, S(x,K) \r)^2}\,,
\end{eqnarray}
where $\int_{\mathrm{bin}} d^3K \l( \cdots \r)$ naturally represents an average over a given $\vec{K}$-bin.  In this paper, all correlation functions are first evaluated as functions of $\vec{q}$ and $\vec{K}$ and then averaged over the azimuthal direction of $\vec{K}$ in the transverse plane, denoted $\Phi_K$.  Only such $\Phi_K$-averaged correlation functions are finally fit to extract, e.g., the L\'evy exponent $\alpha$.

\subsection{One-dimensional projection}

Another way to improve statistical precision is to measure the correlation function as a function of a single scalar quantity which is commonly chosen to be either Lorentz invariant or simply boost invariant in the longitudinal direction \cite{Adare:2017vig}.  This approach has the benefit of reducing statistical errors and improving the precision of the correlation function measurement, but it tends to distort the measured correlation function from the expected Gaussian form when the fully differential correlation function deviates from a spherically symmetric form.

To see how to obtain the projected (i.e., one-dimensional) correlation function from the full, three-dimensional correlation function, let us focus on the numerator of \eqref{correl_func_FT_S}; the denominator can be handled in a similar fashion.  First, we introduce the shorthand
\begin{equation}
	\mathrm{Num}(q,K) \equiv \l| \int d^4 x\, e^{i q \cdot x} S(x,K) \r|^2\,.
\end{equation}
The total number of pairs entering the numerator (at fixed $K$) is independent of whether the correlation function is projected and is obtained, up to an overall normalization factor \cite{Miskowiec:1997ay,Wiedemann:1999qn}, by integrating over $q$:
\begin{equation}
	N_{\mathrm{pair,num}}(K) \equiv \int d^3q\, \mathrm{Num}(q,K)
\end{equation}
This will be unchanged if we insert a factor which is equal to 1 by definition:
\begin{multline}
	N_{\mathrm{pair,num}}(K) \\
	\equiv \int d^3q\, \int \frac{dQ}{2\l| Q \r|} \delta\l( Q^2 - {Q'}^2(q) \r) \mathrm{Num}(q,K),
\end{multline}
where $Q' \in \l( -\infty, \infty \r)$ is a scalar constructed from  the relative momentum $\vec{q}$ which can be defined in different ways.  The Lorentz invariant definition is given by
\begin{equation}
	Q_{\mathrm{inv}}^2 = -q^\mu q_\mu = \vec{q} \cdot \vec{q} - \l( q^0 \r)^2\,. 
	\label{Q_LI}
\end{equation}
This has the disadvantage that $Q_{\mathrm{inv}}^2$ may become 0 even is none of the components of $q^\mu$ vanishes. This is inconvenient, because a peak of the correlation function is expected at vanishing momentum difference, but this feature could reduce the peak. Therefore, in   \cite{Adare:2017vig} a different variable has been proposed, which is not 0 unless all components of $q^\mu$ vanish
\begin{equation}
	Q_{\mathrm{LCMS}}^2 = q_o^2 + q_s^2 + \frac{(p_{1,l}E_2-p_{2,l}E_1)^2}{K_0^2 - K_l^2}\,  .
	\label{Q_BI}
\end{equation}
This definition makes $Q_{\mathrm{LCMS}}^2$ longitudinally boost invariant, while it collapses to a simple interpretation $Q_{\mathrm{LCMS}}^2 = \vec q\cdot \vec q$ in the longitudinally co-moving system, where $K_l=0$.

For any choice of $Q'$, the projected numerator is then constructed by defining
\begin{multline}
	N_{\mathrm{pair,num}}(Q,K) \\
	\equiv \frac{1}{2\l| Q \r|}\int d^3q\, \delta\l( Q^2 - {Q'}^2(q) \r) \mathrm{Num}(q,K)\,.
\end{multline}
The argument proceeds similarly for the denominator, so that the projected correlation function then reads\footnote{Notice that if we had not used the smoothness approximation, the denominator of \eqref{projected_C} would no longer factorize as it does here.}
\begin{multline}
	C( Q, \vec{K} )
		 \approx  1 +\\
			\frac{\int d^3q\, \delta\l( Q^2 - {Q'}^2(q) \r) \l| \int d^4x\, e^{i q \cdot x} S(x,K) \r|^2 }
			{ \int d^3q\, \delta\l( Q^2 - {Q'}^2(q) \r)\l( \int d^4x\, S(x,K) \r)^2}\,. 
		\label{projected_C}
\end{multline}
Customarily, the choice of $Q'$ in the $\delta$-function is indicated by $C(Q_{\mathrm{inv}},K)$ or $C(Q_{\mathrm{LCMS}},K)$.  We will consider both possibilities
in this study.

\subsection{Resonance decays}
The effects of resonance decays on the HBT radii have been studied previously \cite{Wiedemann:1996ig} and are by now well known.  Each resonance which contributes to the final pion yield must first decay into lighter particles; the typical timescale on which this happens in the resonance's rest frame is set by its inverse width.
This introduces a corresponding scale in the total pion emission function which represents the typical distance a resonance propagates before decaying.

Moreover, since many different resonances contribute to the final pion yield, many different lengthscales and timescales are therefore represented in the space-time structure of the total pion emission function.  It is a fundamental property of Gaussian distributions that they may be fully characterized in terms of a single lengthscale, namely, their standard deviation $\sigma$; conversely, any distribution characterized by multiple lengthscales cannot be Gaussian in form.  Hence, since the total pion emission function contains multiple different lengthscales, it must therefore deviate from a Gaussian form once resonance effects are taken into account.

This last observation is true, even for the fully differential correlation function, meaning that the Gaussian form fails to capture the complete structure of \eqref{correl_func_def}.  Fortunately, the resulting non-Gaussianities can still be quantified and studied systematically using the ``$q$-moments" proposed in \cite{Wiedemann:1996ig}.  For most resonances, the effects of resonance decays are mainly confined to a narrow region about the origin $\vec{q}=0$, and the space-time structure of pion emission directly from the freeze-out surface can be isolated by focusing the analysis at somewhat larger values of $\l| \vec{q} \r|$ \cite{Frodermann:2006sp,Plumberg:2015mxa}.  

\section{Model Studies}
\label{ModelStudies}

Having presented several different effects which can lead to non-Gaussianities in the measured correlation function, we now proceed to quantify the magnitude of each effect on a typical correlation function.  In particular, the modeling of resonance decays requires the ability to simulate the effects of the decay cascade, for which purpose we consider two different packages for modeling heavy-ion collisions.  In this section, we detail their basic elements and structure, and in the next section we consider the effects of each source of non-Gaussianity within the context of each model.

\vspace{-5.5pt}
\subsection{The blast-wave model}
\label{BlastWaveModel}

The blast-wave model  \cite{Siemens:1978pb,Schnedermann:1993ws,Csorgo:1995bi,Tomasik:1999cq,Retiere:2003kf} is a parametrisation of emission from a thermalized fireball, which expands longitudinally in a boost-invariant way and also transversely.  Our version of the model allows for azimuthal anisotropies in the shape as well as the transverse expansion. It is defined by its emission function
\newpage
\begin{widetext}
\begin{equation}
S(x,p) \, d^4x = \frac{1}{(2\pi)^3}\,\left (\exp\left ( \frac{u^\mu(x) p_\mu}{T}\right ) \pm 1 \right )^{-1}
\Theta(R(\theta)-r) \, \delta(\tau - \tau_{fo})\, m_t \cosh(\eta-y) \tau\, d\tau\, d\eta\, r\, dr\, d\theta\,  .
\end{equation}
Due to existing symmetries and the dominant dynamics, radial coordinates $r$ and $\theta$ are used in the transverse plane, and the other two coordinates are space-time rapidity $\eta$ and longitudinal proper time $\tau$ 
\begin{equation}
\eta = \frac{1}{2} \ln \frac{t+z}{t-z}\,  , \qquad \tau = \sqrt{t^2 - z^2}\,   .
\end{equation}
To express the four-momentum of the particle we use its transverse mass $m_t$ and rapidity $y$.

In the transverse direction the fireball has a sharp cutoff set by $\Theta(R(\theta)-r)$, where the transverse radius oscillates in azimuthal angle
\begin{equation}
R(\theta) = R_0 \left [  1 - a_2 \cos\left ( 2 ( \theta - \theta_2 ) \right ) \right ]\,  .
\end{equation}
Here, $R_0$ is the (mean) transverse radius, $a_2$ is the amplitude of the second-order oscillation, and $\theta_2$ is the direction of the second-order event plane. All freeze-out happens along the hypersurface set by longitudinal proper time $\tau_{fo}$ and  $m_t \cosh(\eta-y) \tau\, d\tau\, d\eta\, r\, dr\, d\theta$ is the Cooper-Frye factor which stands for the flux of the particles through this hypersurface \cite{Cooper:1974mv}.

Thermal production is encoded in the standard Fermi-Dirac or Bose-Einstein distribution, where $u^\mu(x) p_\mu$  gives the energy of the particle in the local rest frame of the fluid. The expansion is expressed via the velocity field 
\begin{equation}
u^\mu(x) = (\cosh\eta\, \cosh\eta_t(r,\theta_b),\, \cos\theta_b\, \sinh\eta_t(r,\theta_b), \\  \sin\theta_b\, \sinh\eta_t(r,\theta_b),\, 
\sinh\eta\, \cosh\eta_t(r,\theta_b) )\,   ,
\end{equation}
\end{widetext}
The transverse velocity is here parametrised with the help of transverse rapidity $\eta_t$ 
\begin{equation}
v_t = \tanh\eta_t(r,\theta_b)\,  ,
\end{equation}
which grows with the distance from the longitudinal symmetry axis of the fireball and shows also a second-order variation
\begin{equation}
\eta_t (r,\theta_b) = \rho_0\frac{r}{R(\theta)} \left [ 1 + 2\rho_2 \cos(2(\theta_b - \theta_2)) \right ]\,  ,
\end{equation}
where $\rho_0$ and $\rho_2$ are parameters of the model. Note that the transverse rapidity varies with $\theta_b$ and not with $\theta$; the former is directed always so that it points perpendicularly to the surface with constant $r/R(\theta)$ and is given as 
\begin{equation}
\tan \left ( \theta_b - \frac{\pi}{2}\right )
= \frac{dx_2}{dx_1} = \frac{\frac{dx_2}{d\theta}}{\frac{dx_1}{d\theta}} = \frac{  \frac{d R(\theta)\sin(\theta)}{d\theta}}{  \frac{d R(\theta)\cos(\theta)}{d\theta}} \,  ,
\end{equation}
where the functional dependences $x_1(\theta)$, $x_2(\theta)$ refer to the transverse boundary of the fireball \cite{Cimerman:2017lmm}.

For the actual calculation of various effects we will generate artificial events with the help of DRAGON Monte Carlo event generator \cite{Tomasik:2008fq,Tomasik:2016skq}, which is based on this blast-wave model. In addition to that, DRAGON also includes the production and decay of resonances. They are produced from the same emission function with their pole masses, and they decay exponentially in their own time according to their widths. Both two- and three-body decays are included as well as the possibility that one resonance type can decay via various channels according to their branching ratios. Cascades of decays, in which several resonances decay consecutively, are also possible within the model.

For this study, we used DRAGON to generate sets of 50,000 events. The basic setting of the parameters used includes the temperature of 120~MeV, the average transverse radius $R=7$~fm, freeze-out time $\tau_{fo} = 10$~fm/$c$, and the strength of the transverse expansion $\rho_0 = 0.8$.

The correlation functions from the simulated samples are generated by CRAB \cite{CRAB} and fitted to find the best parametrisation. Hence, this procedure resembles the one used by experimentalists and the correlation function is obtained in bins in the momentum difference $q$.

\subsection{Hydrodynamics}
\label{Hydrodynamics}

In addition to studying the correlation functions obtained from the blast-wave model, we also consider an approach based on hydrodynamical modeling of the collision system using the iEBE-VISHNU model \cite{Song:2007ux,Shen:2014vra}.  This model is then extended to HBT using the \HoTCoffeeh (\textbf{H}anbury Br\textbf{O}wn-\textbf{T}wiss \textbf{CO}rrelation \textbf{F}unctions and radii \textbf{F}rom \textbf{E}vent-by-\textbf{E}vent \textbf{H}ydrodynamics) package \cite{Plumberg:2016sig} which computes the event-by-event correlation functions entirely in terms of Cooper-Frye integrals. 
Note that this determination of the correlation function is thus in principle exact.
We briefly describe the most important elements of these modeling tools for our purposes; further details can be obtained in the relevant references \cite{Shen:2014vra,Plumberg:2016sig}.

The iEBE-VISHNU package \cite{Shen:2014vra} combines a Monte-Carlo initial state model (e.g., MC-Glauber \cite{Broniowski:2007nz,Alver:2008aq,Loizides:2014vua}) with hydrodynamic evolution which is determined by the boost-invariant Israel-Stewart equations \cite{Israel:1979wp,Song:2007ux}.  The iEBE-VISHNU package is well-documented and publicly available for download \cite{iEBEVISHNUdownload}.  The $N_{\mathrm{ev}} = 1000$ events used in this study were generated for 0-10\% Au+Au collisions at 200$A$ GeV using MC-Glauber hydrodynamic initial conditions, a value of $\eta/s = 0.08$ during the hydrodynamic evolution, and a freeze-out temperature of $T_{fo} = 120$ MeV.

The \HoTCoffeeh code \cite{Plumberg:2016sig} was recently developed to compute the HBT correlation function on an event-by-event basis by directly evaluating Cooper-Frye \cite{Cooper:1974mv} integrals over the freeze-out surface.  The freeze-out integrals themselves contain contributions both from directly (`thermally') produced particles and also from resonance decays.  In principle,  \HoTCoffeeh is capable of evaluating all contributions from resonances in the PDG book \cite{Beringer:1900zz,myPDGfile} up to and including the mass of the $\Omega(2250)^{-}$ baryon.  This corresponds to roughly 320 resonances and 1500 decay channels, many of which contribute only negligibly to the final pion yields.

In order to accelerate the computation of the correlation function with resonance effects included, \HoTCoffeeh evaluates only the small subset of all resonances which is necessary to reach a given threshold of the total decay contribution (e.g., in this study, we considered only the most important resonances necessary to produce 60\% of the total pion yield coming from resonance decays).  Once the threshold is reached, \HoTCoffeeh extrapolates linearly from the partial resonance yield to estimate the total resonance contribution.  This trick, which was first introduced in \cite{Qiu:2012tm} for approximately evaluating resonance contributions to the anisotropic flow coefficients $v_n$, reduces the average runtime of the code by several orders of magnitude, and was shown in \cite{Plumberg:2016sig} to produce reasonable agreement with the full resonance calculation.

\section{Results}
\label{Results}

\begin{figure}[t]
\includegraphics[width=\linewidth]{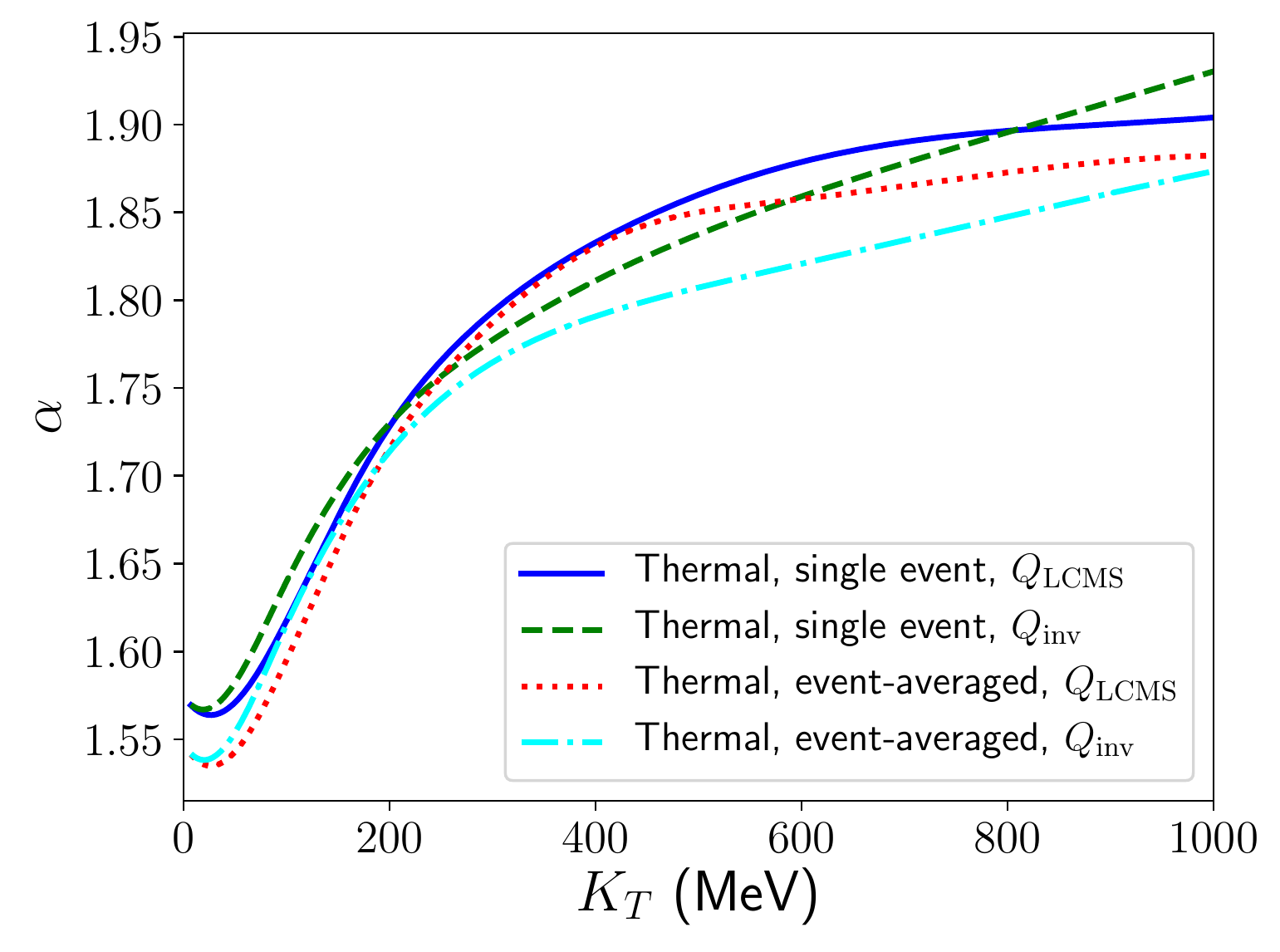}
\includegraphics[width=\linewidth]{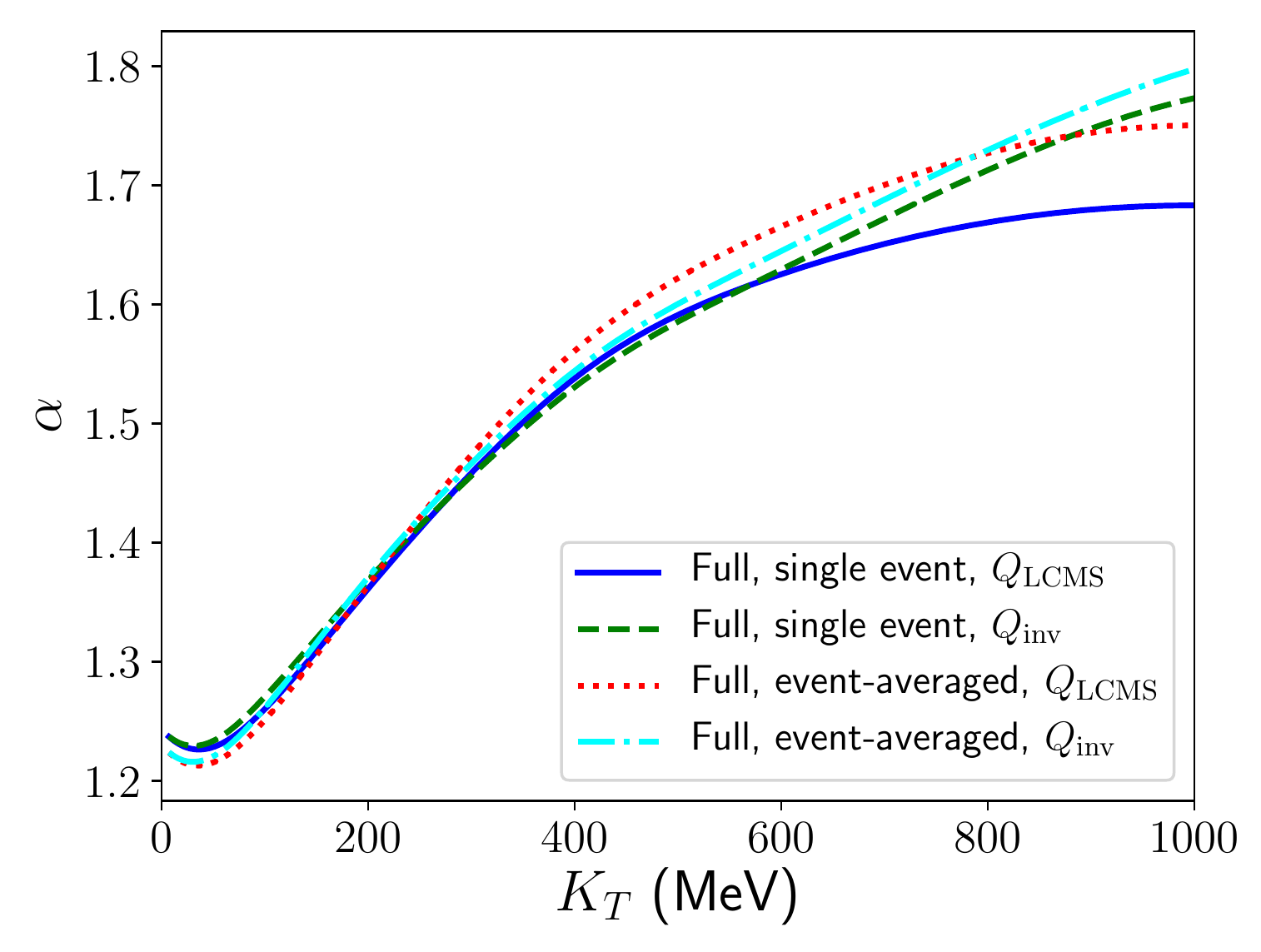}

\caption{A comparison of $\alpha(K_T)$ with and without different non-Gaussian effects: with and without ensemble averaging (solid blue and dashed green vs. dotted red and dash-dotted cyan) and one-dimensional projection in hydrodynamics, for different choices of $Q$, corresponding either to Eq.~\eqref{Q_LI} or to Eq.~\eqref{Q_BI}.  The comparison is made both for thermal pions only (upper panel) and for the full thermal and resonance contributions added together (lower panel).  Despite small quantitative differences, comparison of these curves shows that neither the choice of $Q$ nor the inclusion of ensemble averaging has a dramatic effect on the $K_T$-dependence of $\alpha$.
\label{f:LI_vs_BI}}
\end{figure}

We begin our model studies by testing the sensitivity of our results to the definition used for $Q$, given either by Eq.~\eqref{Q_LI} or by Eq.~\eqref{Q_BI}.  This comparison is made using the hydrodynamic approach, with the results shown in Fig.~\ref{f:LI_vs_BI}.
The differences between $Q_{\mathrm{inv}}$ and $Q_{LCMS}$ can be appreciated by comparing the blue solid and red dotted curves with the corresponding green dashed and cyan dash-dotted curves in Fig.~\ref{f:LI_vs_BI}.  From these comparisons, we find that the non-Gaussian sources of event averaging and the choice of $Q_{\mathrm{inv}}$ or $Q_{LCMS}$ matters only at the quantitative level for the shape of $\alpha(K_T)$; the qualitative trends are mostly unaffected by these choices.  

Fig.~\ref{f:LI_vs_BI} also illustrates the effects of ensemble averaging on the shape of the measured correlation function.  The single-event (SE) curves in Fig.~\ref{f:LI_vs_BI} are represented by the blue solid and green dashed curves, while the ensemble-averaged (EA) curves are indicated by the red dotted and cyan dash-dotted curves.  One finds that, for the one-dimensional projections of the correlation function, the various $\alpha(K_T)$ curves show surprisingly little sensitivity to the effects of event-by-event fluctuations.  Hereafter, unless stated otherwise, we present only ensemble-averaged results.

Similar conclusions regarding the importance of ensemble averaging can be obtained from the blast-wave approach.  We begin by considering the effects that the averaging over many fireballs with different shapes may have on the value of the L\'evy parameter $\alpha$. Indeed, in real experiments each fireball is different, with different sizes, eccentricities, and orientations of the event plane.  We therefore anticipate that averaging over a distribution of source shapes, as is implied by \eqref{avg_correl_func_FT_S}, will cause $\alpha$ to deviate from 2.  In this treatment, we limit our focus to second-order anisotropies.

We first study the effect of averaging over different values of the spatial anisotropy parameter $a_2$. 
%
\begin{figure*}[t]
\centering
\includegraphics[width=0.32\textwidth]{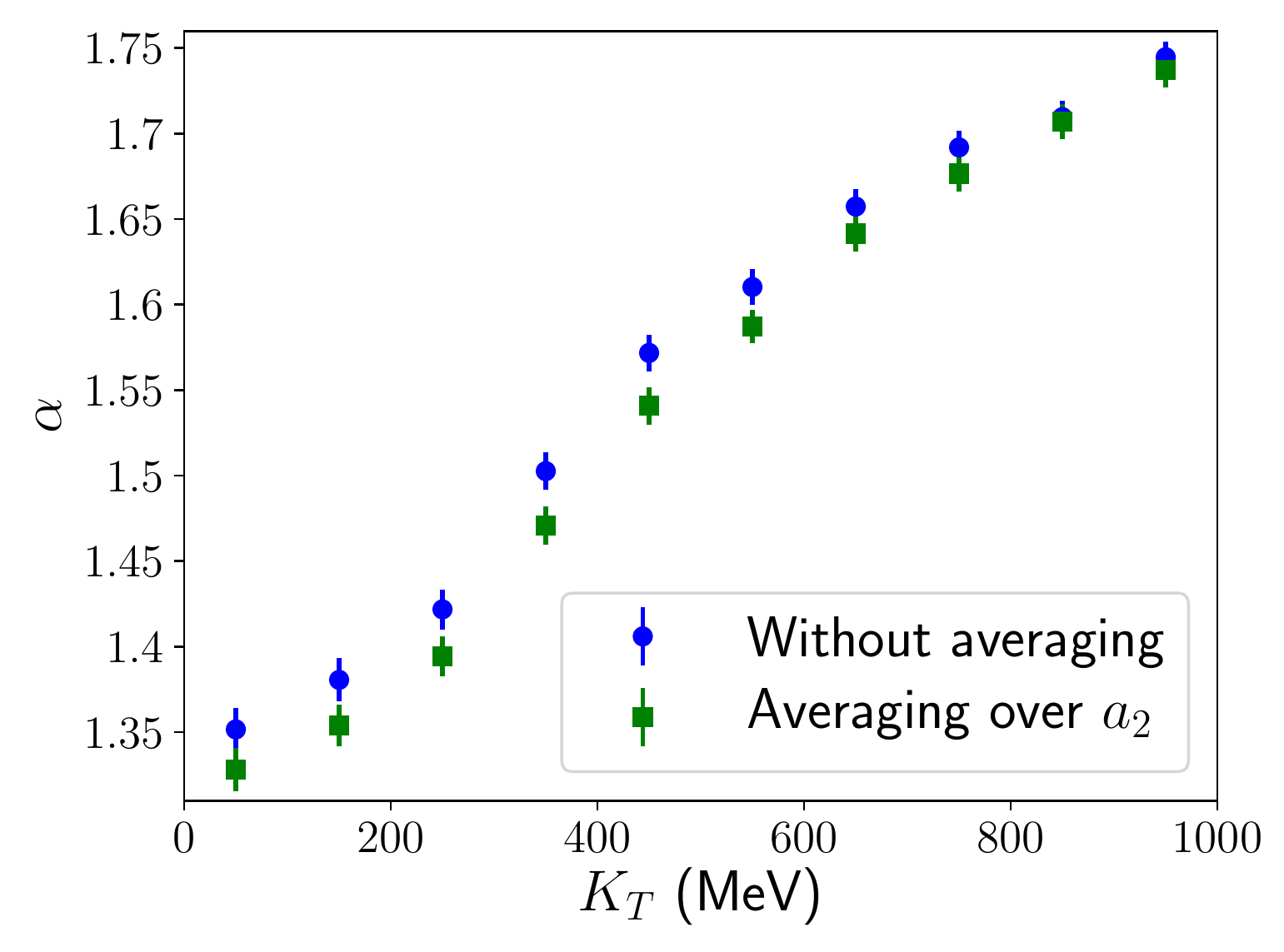}
\includegraphics[width=0.32\textwidth]{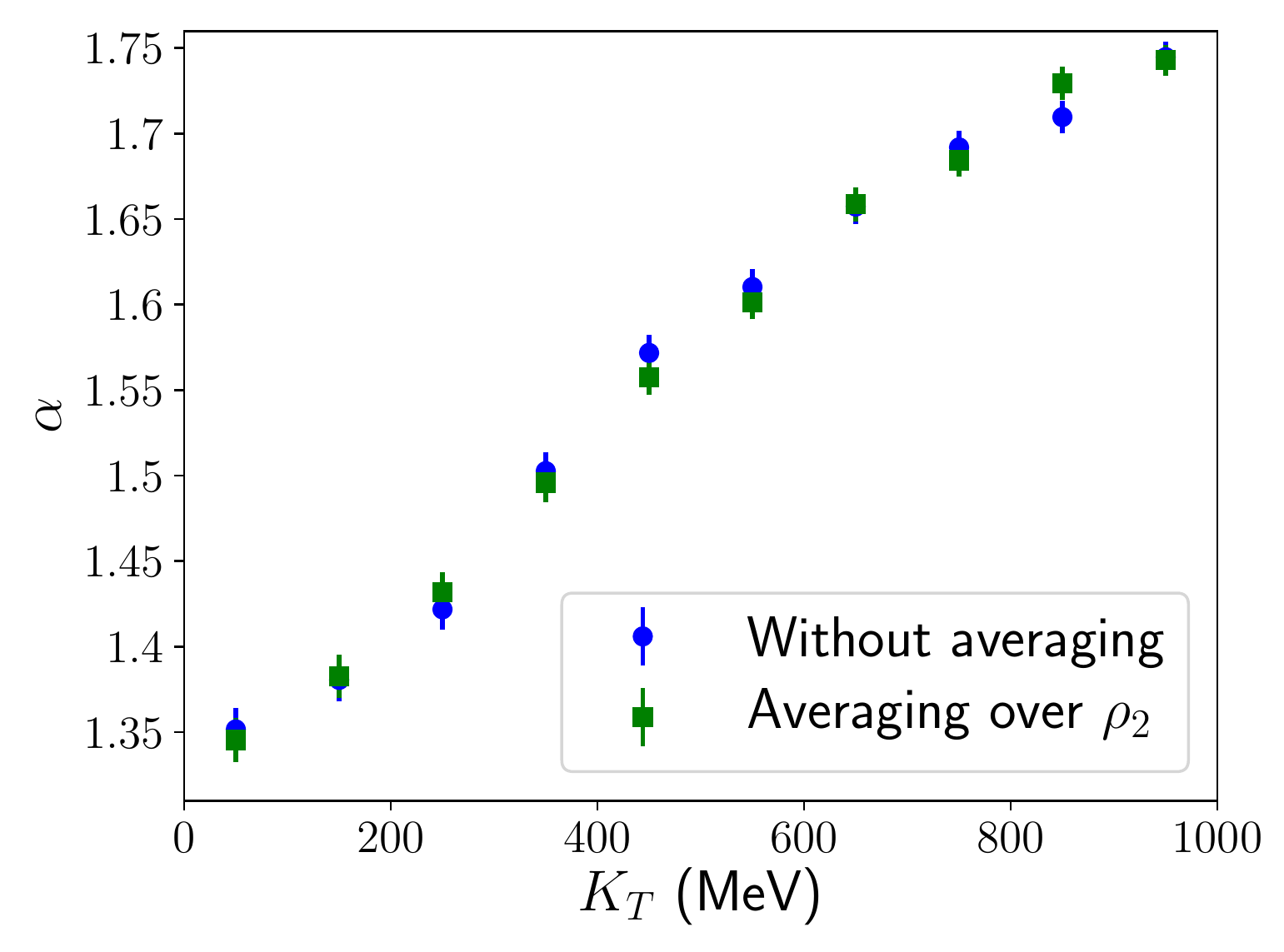}
\includegraphics[width=0.32\textwidth]{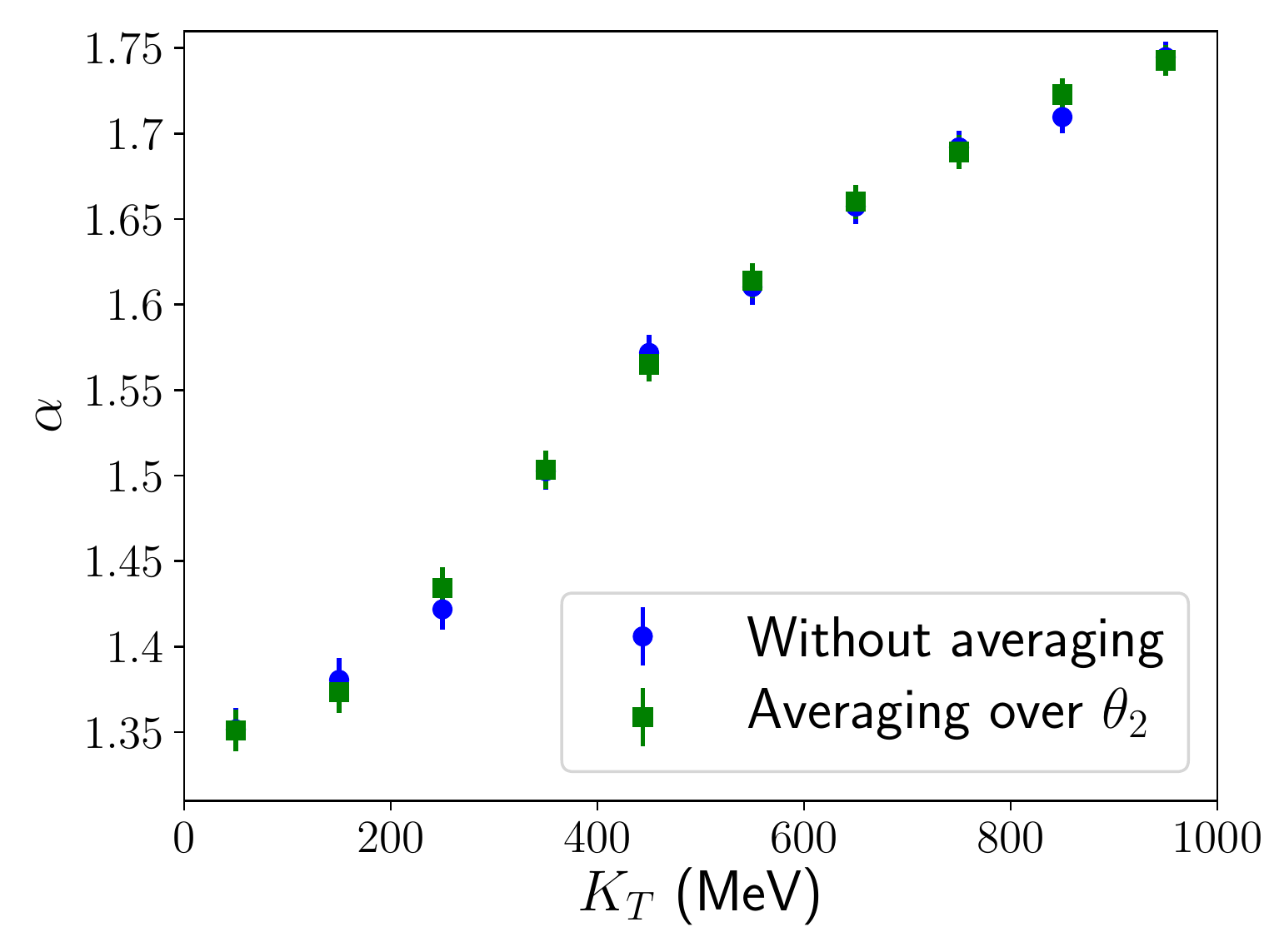}
\caption{The L\'evy parameter of the 1D fit to the correlation function in $Q_{\mathrm{LCMS}}$. Mean transverse momentum $K_T$ in bins of 100~MeV. The green points show results calculated with fixed anisotropies, while the blue points show results calculated for averaging over $a_2$ (left); averaging over $\rho_2$ (middle); and averaging over $\theta_2$ (right).
\label{f:eaver}}
\end{figure*}   
%
Figure \ref{f:eaver} (left) compares the L\'evy parameter $\alpha$ from two sets of Monte Carlo events. In the first set, all events have the spatial eccentricity with $a_2 = 0.05$. In another set, the eccentricity fluctuates with $a_2$ between $-0.1$ and $0.1$. 
We see that the values of $\alpha$ departs from 2 considerably and reaches values between 1.27 and 1.75. The averaging makes up only a small portion of this decrease, at most at a level of 0.05. 

Almost identical results quantitatively come from the averaging over flow anisotropy (Fig.~\ref{f:eaver}, middle) and the event plane orientation (Fig.~\ref{f:eaver}, right). In the middle panel we compare L\'evy index $\alpha$ obtained from a set with $\rho_2$ fixed to 0.05 with a set with events for which $\rho_2$ fluctuates between $-0.1$ and $0.1$. For the event plane averaging we see no change 
if $\theta_2$ fluctuates in comparison to $\theta_2$ fixed to 0. The anisotropy parameters $a_2$ and $\rho_2$ in this case fluctuate between $-0.1$ and $0.1$.

We investigate next the influence of resonances on the obtained value of $\alpha$. We use the basic source of particles in the blast-wave model with the same basic parameters as in the previous case, and we compare correlation functions obtained with and without resonance decays.  We also include corresponding calculations resulting from hydrodynamics, using the set-up described in the previous section.
%
\begin{figure}[t]
\centering
\includegraphics[width=\linewidth]{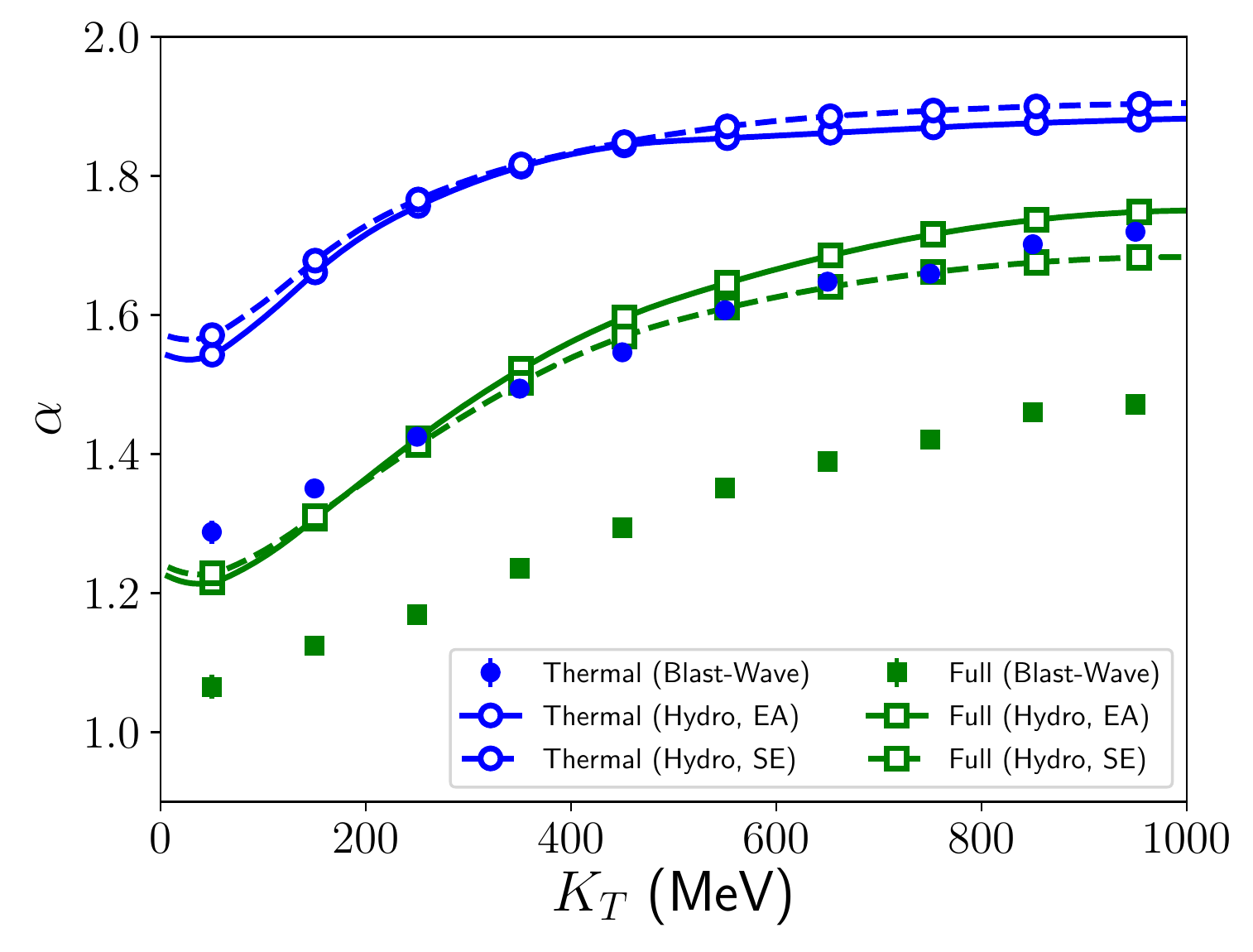}
\caption{\label{f:res}The L\'evy parameter $\alpha$ of the 1D fit to the correlation function in $Q_{\mathrm{LCMS}}$. Result from the fits to correlation function from a source without resonances (blue circles) and with resonances (green squares).  The solid points with error bars correspond to the BW approach, while the open points connected by lines represent the hydrodynamic results, both for ensemble-averaged (EA, solid) and for single-event (SE, dashed) correlation functions.}
\end{figure}   
%
In Figure~\ref{f:res} we show the L\'evy indices $\alpha$ obtained from fits to the correlation functions as a function of $K_T$. We know from the previous Figure already, that the one-dimensional correlation function has quite a non-Gaussian shape. Now we see that the inclusion of the resonance decays pushes down the value of $\alpha$ by another 0.2. The influence of resonance decays is much bigger than that of averaging over different events!  This clearly holds whether one uses the BW model or hydrodynamics to describe the collision evolution and freeze-out.  Notice also that the non-Gaussianity is stronger at small than at large $K_T$, an effect which can be at least partially attributed to the dramatic differences between the magnitudes of the longitudinal and transverse radii at small $K_T$.

To gain further insight into these results, we repeat our analysis in a more differential fashion by considering independent (one-dimensional) L\'evy fits along each axis of the full differential correlation function separately.  The fits are performed with the L\'evy prescription \eqref{correl_func_Levy_1D_form}. This is not a three-dimensional analysis, since we do not fit the correlation function in the whole $q$-space; instead, we confine ourselves only to fits along the axes.  The aim is to see the differences in its shape along different directions. 
%
\begin{figure}
\centering
\includegraphics[width=\linewidth]{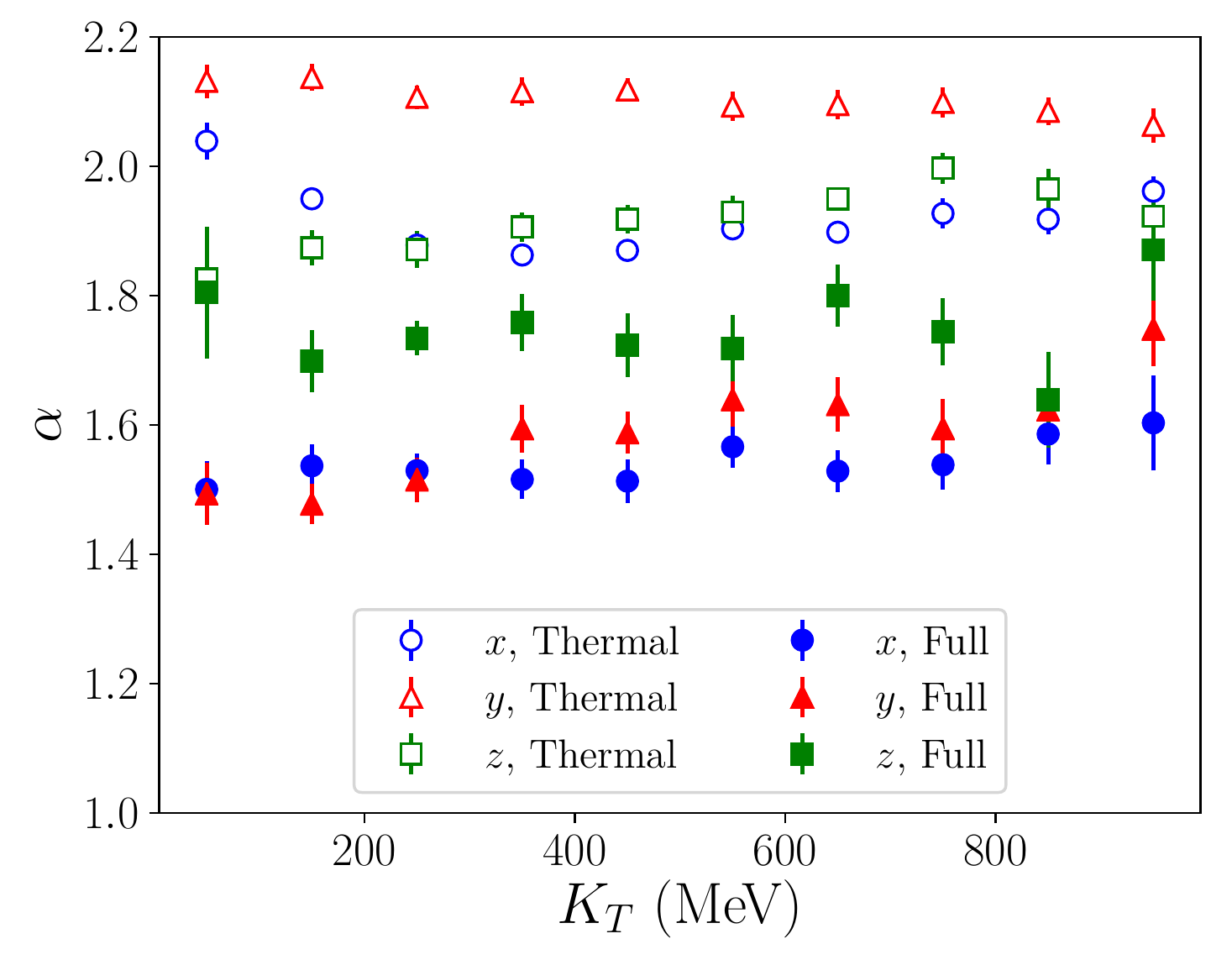}
\includegraphics[width=\linewidth]{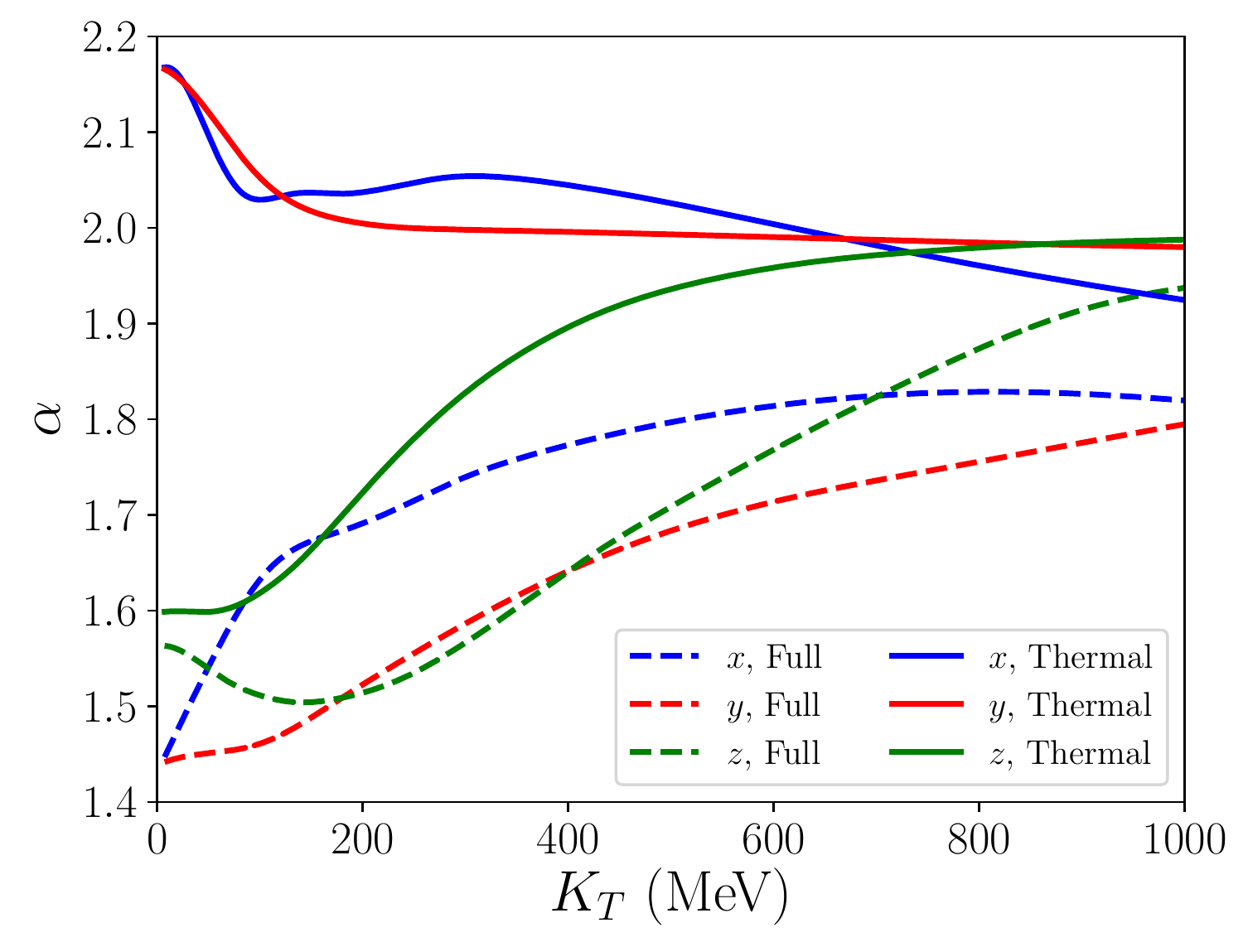}
\caption{The L\'evy parameter $\alpha$ from 1D fits to the correlation function in $Q_{\mathrm{LCMS}}$ along the different axes, with or without resonances.  Upper panel: BW model.  Lower panel: hydrodynamics.
\label{f:dirs}}
\end{figure}   
%
Indeed, we observe in Fig.~\ref{f:dirs} that the differences are rather large. First, consider the blast-wave model results presented in the upper panel.  If resonance decays are not included, the value of $\alpha$ is around 2 in both transverse directions. However, in the longitudinal direction the L\'evy index $\alpha$ is lowered to 1.8 at $K_T = 0$ and increases gradually towards 2 at $K_T  = 1$~GeV. Also, the influence of resonance decays is different for longitudinal and transverse directions. In the longitudinal direction the resonance decays cause a decrease of $\alpha$ by about 0.2. In transverse directions, however, $\alpha$ drops as low as 1.5 due to inclusion of resonance decays. 

These conclusions are reinforced by considering the hydrodynamic approach (lower panel of Fig.~\ref{f:dirs}).  The transverse radii obtained from thermal pions in this case have $\alpha \sim 2$, except for some minor deviations at small $K_T$ and a dip at large $K_T$ in the outward direction.  The longitudinal radius obtained from thermal pions, on the other hand, is not very well described by a Gaussian at small $K_T$, as is already well known \cite{Wiedemann:1995au}.

Once resonance effects are included in the hydrodynamic approach, all three directions deviate much more strongly from a Gaussian form.  The effects are again most pronounced at small $K_T$ where the effects of resonance decays are most important; at large $K_T$, thermal pion production again begins to dominate and the emission source again becomes more Gaussian in form \cite{Plumberg:2016sig}.

We would like to understand these differences further from the perspective of the blast-wave model.   To do this, we directly check the shape of the source which emits pions in this approach.

%
\begin{figure*}[t]
\includegraphics[width=0.32\textwidth]{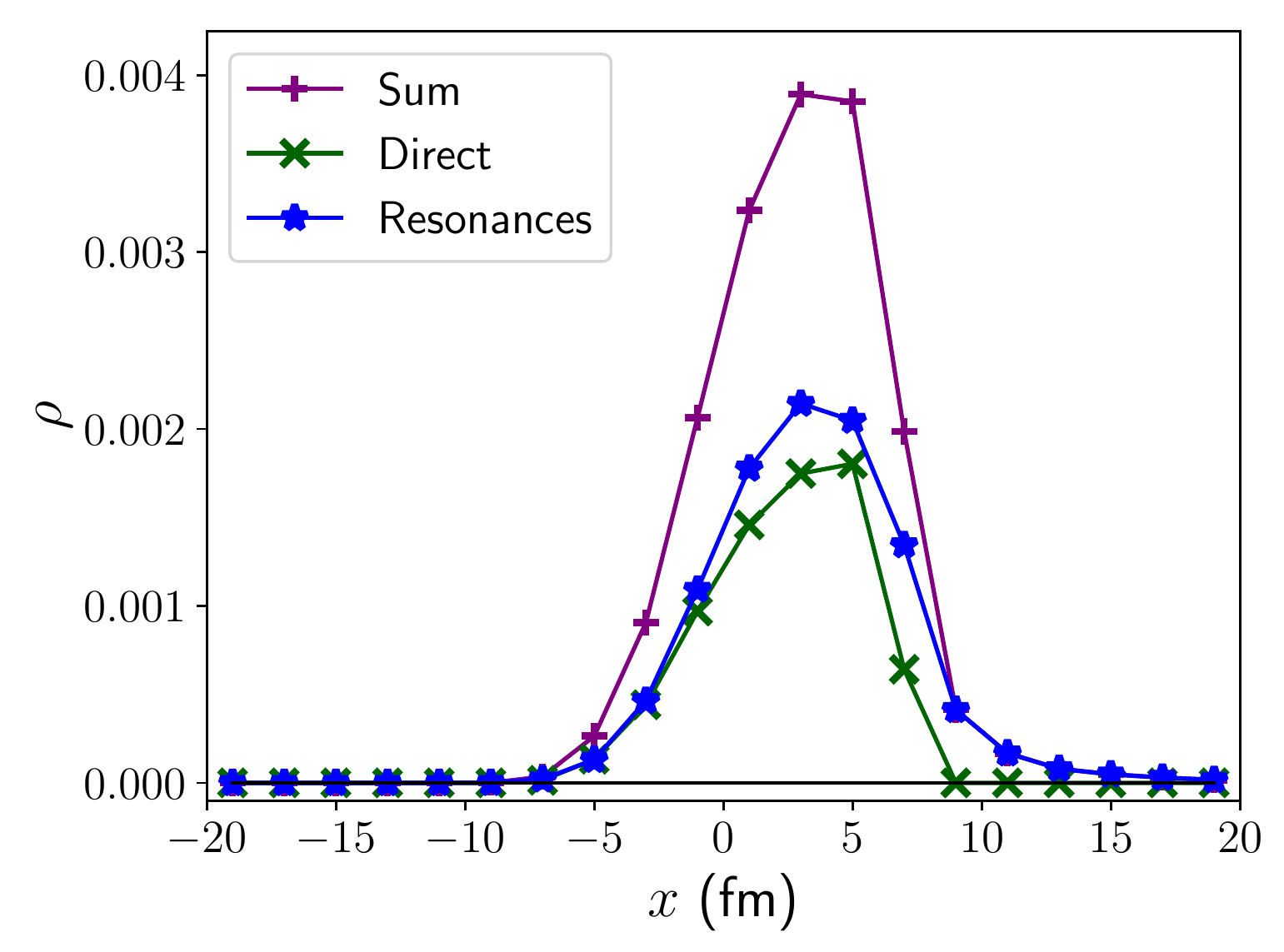}
\includegraphics[width=0.32\textwidth]{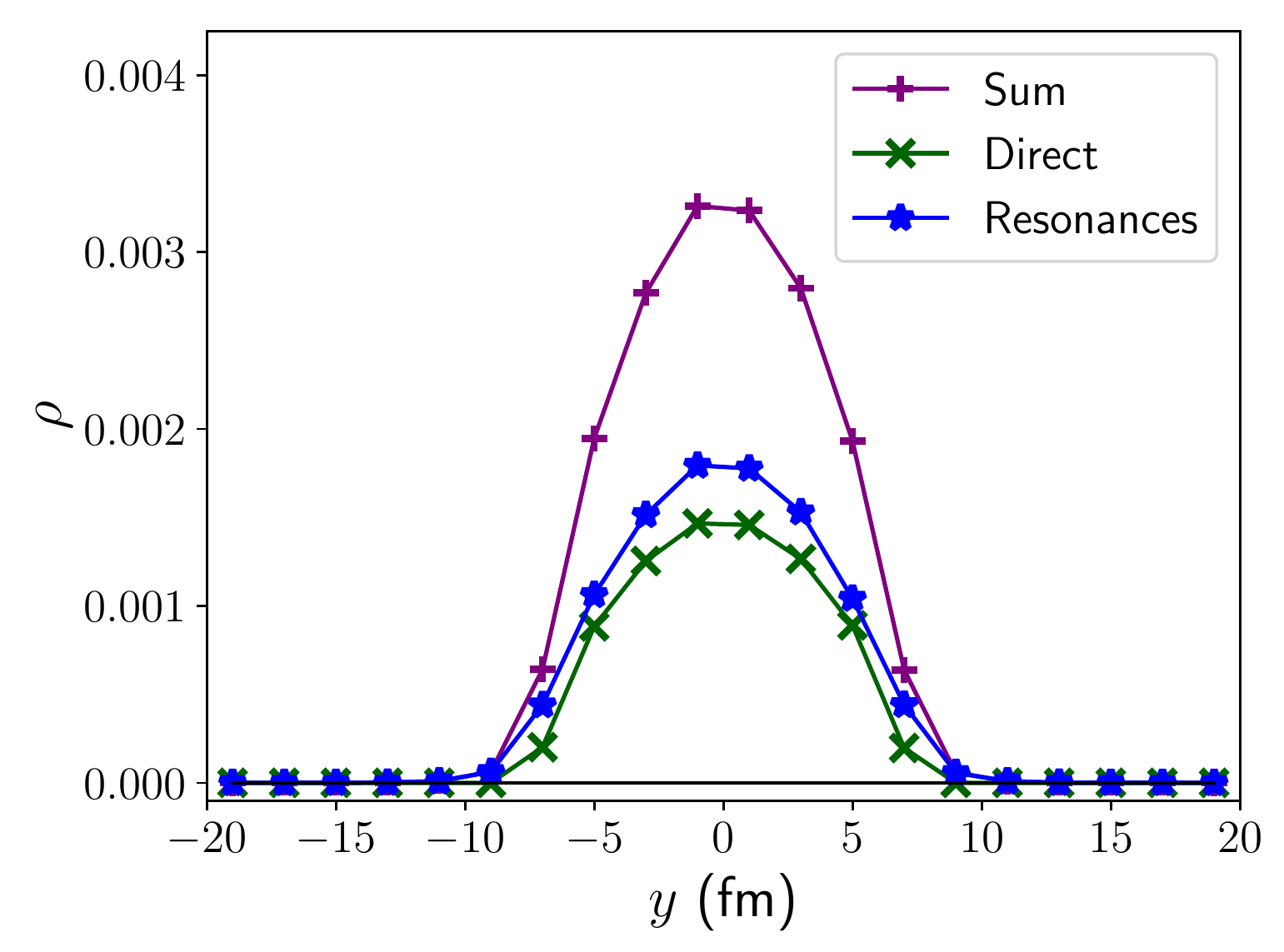}
\includegraphics[width=0.32\textwidth]{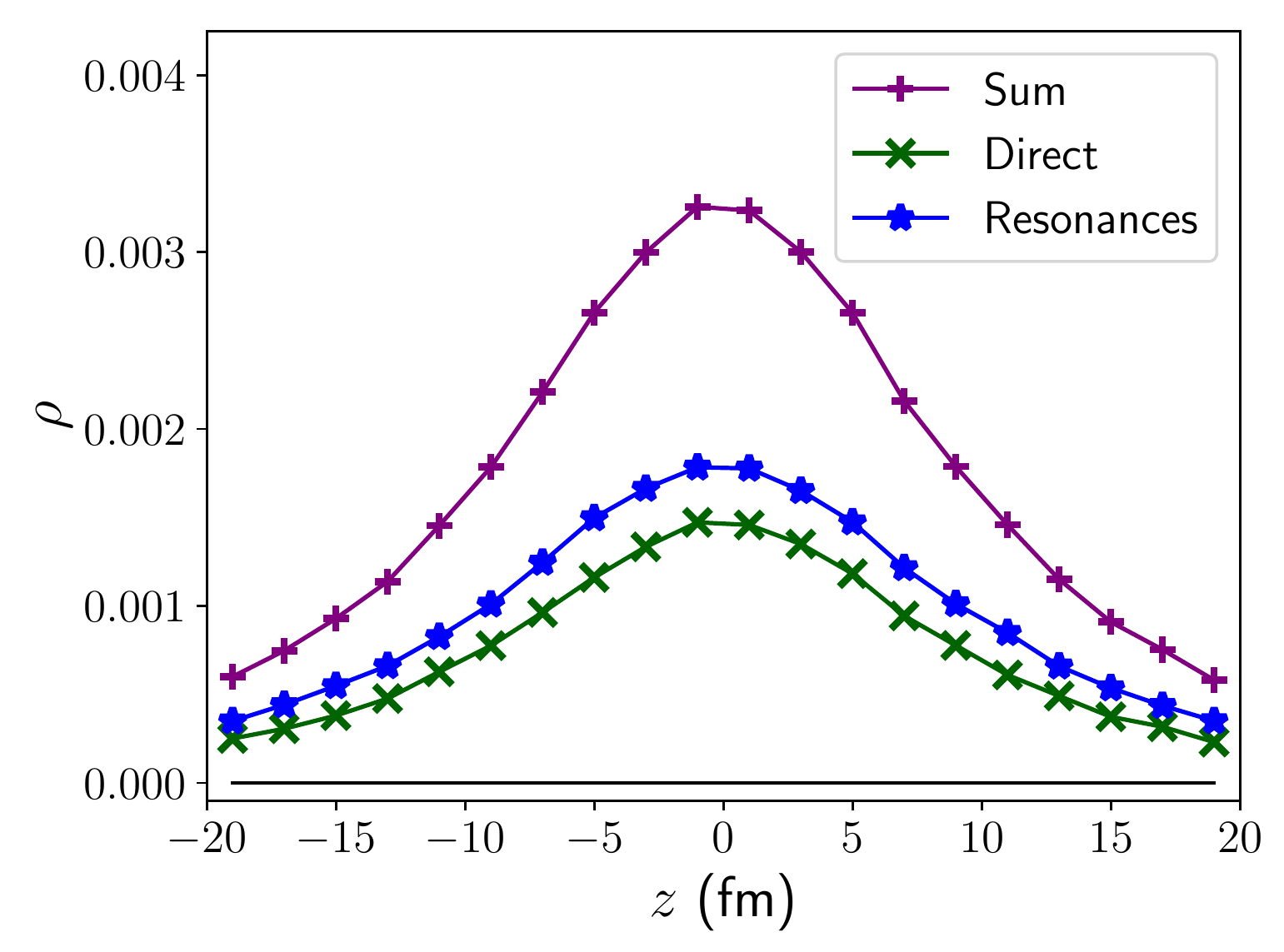}
\vspace*{0.4cm}
\includegraphics[width=0.32\textwidth]{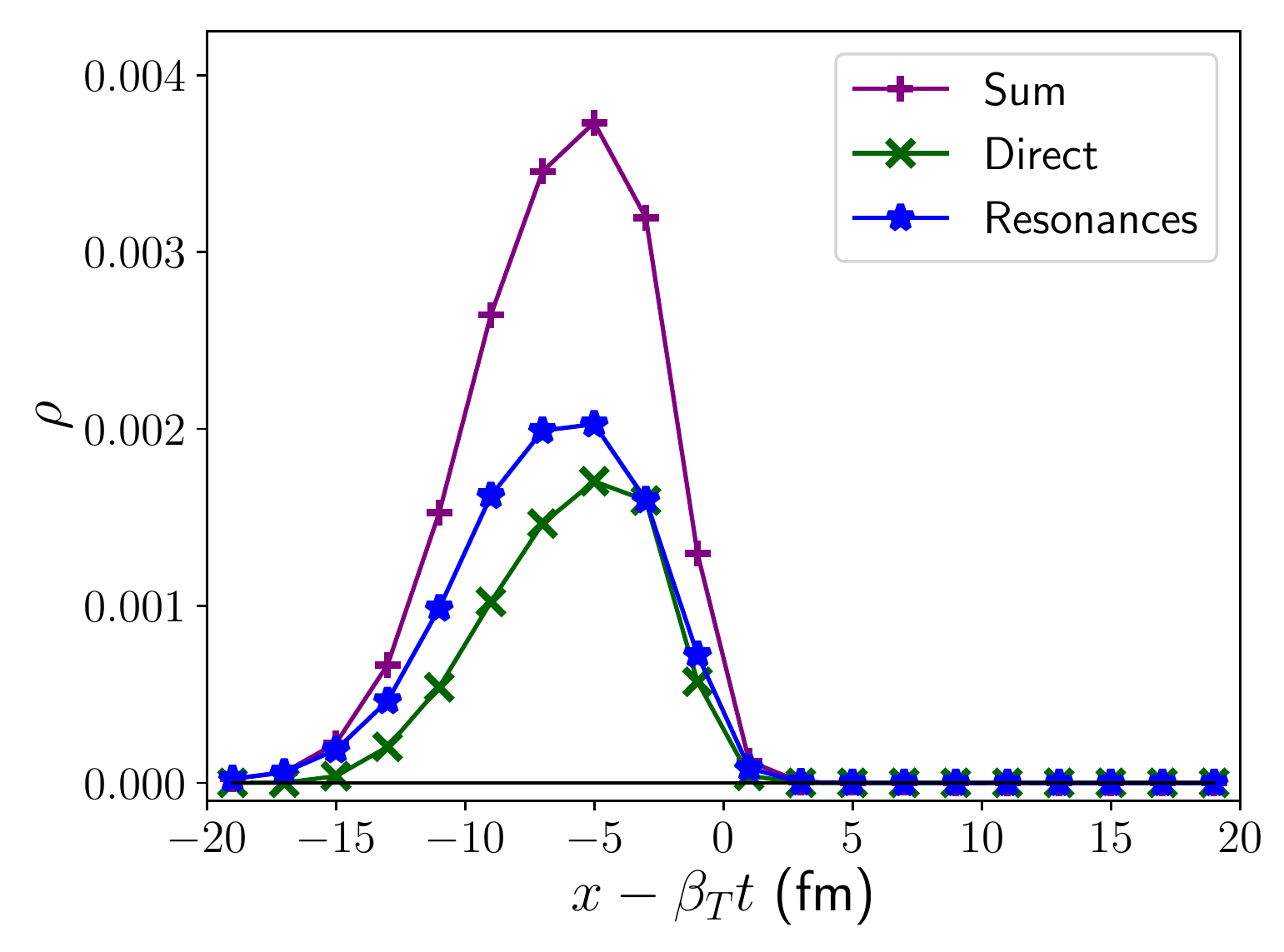}
\includegraphics[width=0.32\textwidth]{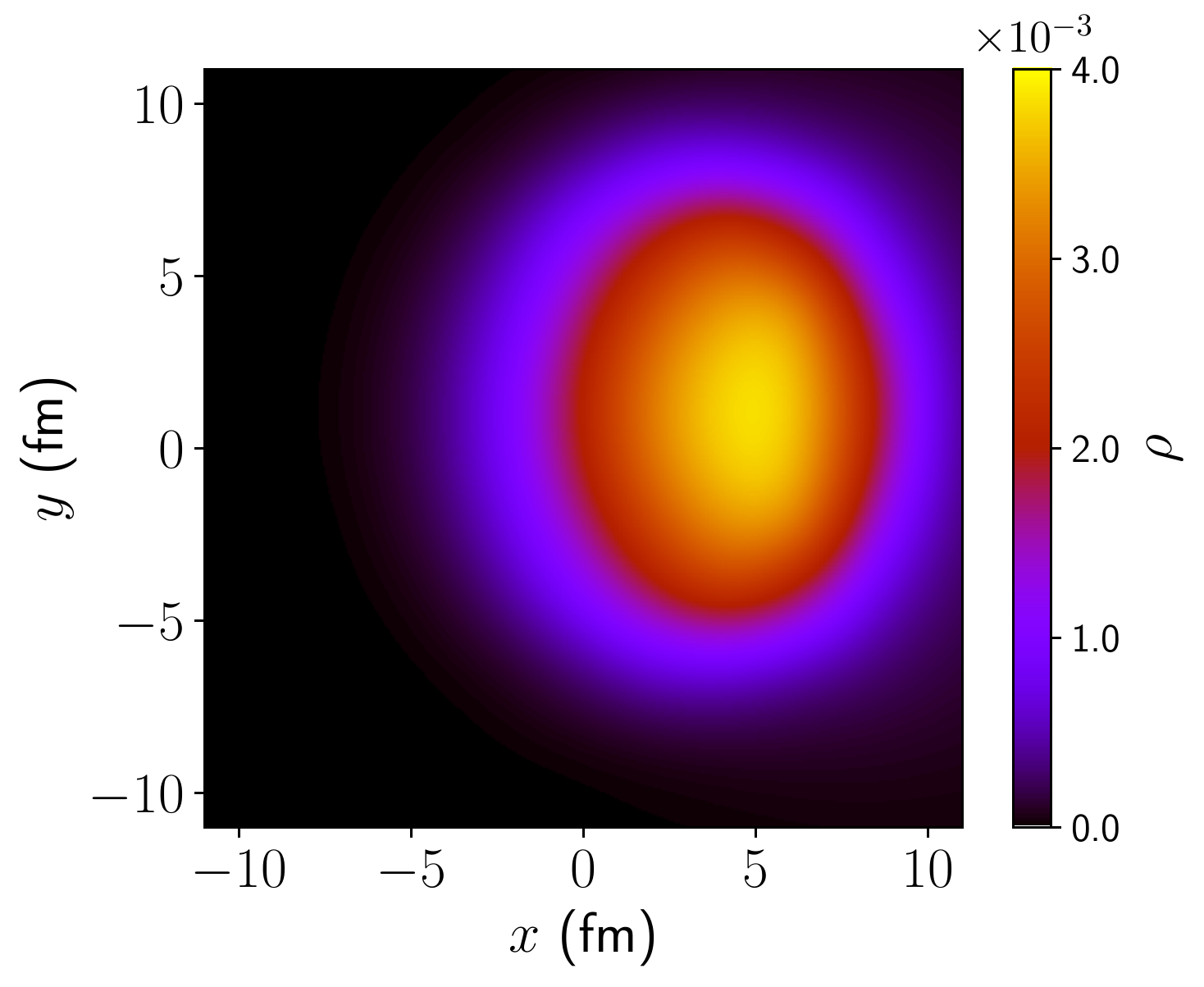}
\includegraphics[width=0.32\textwidth]{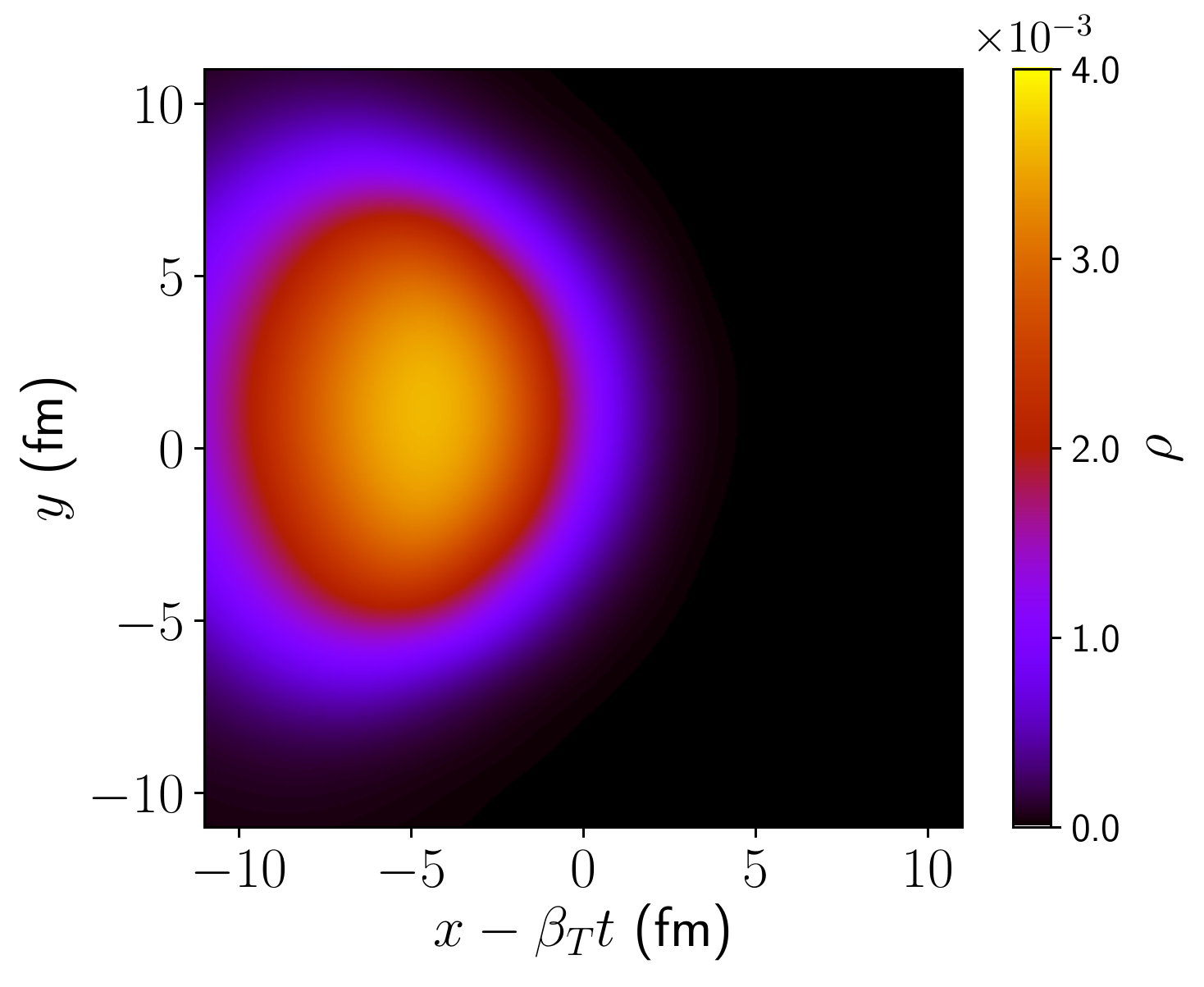}
\caption{The spatial distribution of the emission points of pions. Upper row: the profiles of the emission points distribution along the x (left), y (middle), and z-axis (right). Lower row: the profile along the variable $({\mathrm x} - \beta_t t)$ (left), and two-dimensional distributions in the transverse plane (middle and right). The green $\times$'s show the profile of direct pions, the blue $\ast$'s show the profile of pions produced by resonances and purple $+$'s show their sum. All these distribution were calculated as narrow integrals over the remaining coordinates with width $2$~fm.
\label{f:profiles}}
\end{figure*}   
%

The profiles of the emission function are plotted in Fig.~\ref{f:profiles}, for pions with transverse momentum between 300 and 400 MeV. We show the distribution of the production points of pions, with pions from resonance decays included. The upper row shows that there is quite a difference between the longitudinal and the transverse directions. One could argue, however, that due to the on-shell constraint \eqref{onshell_condition} it is not the distribution in x, that is measured, but rather the distribution in $({\mathrm x} - \beta_t t)$. We plot this in the lower left panel of Fig.~\ref{f:profiles}. In the gradient plots in the two right panels of the lower row we show, 
that both in the $({\mathrm x}, \mathrm{y})$ and $({\mathrm x} - \beta_t t,\mathrm{y})$ planes the source is asymmetric.

Finally, we extend the fitting to the whole three-dimensional correlation function from the previous simulations. The fit is performed with the three-dimensional L\'evy distribution \eqref{correl_func_Levy_3Ddiag_form}, so it always results in a single value of $\alpha$. 
%
\begin{figure}[h!]
\centering
\includegraphics[width=\linewidth]{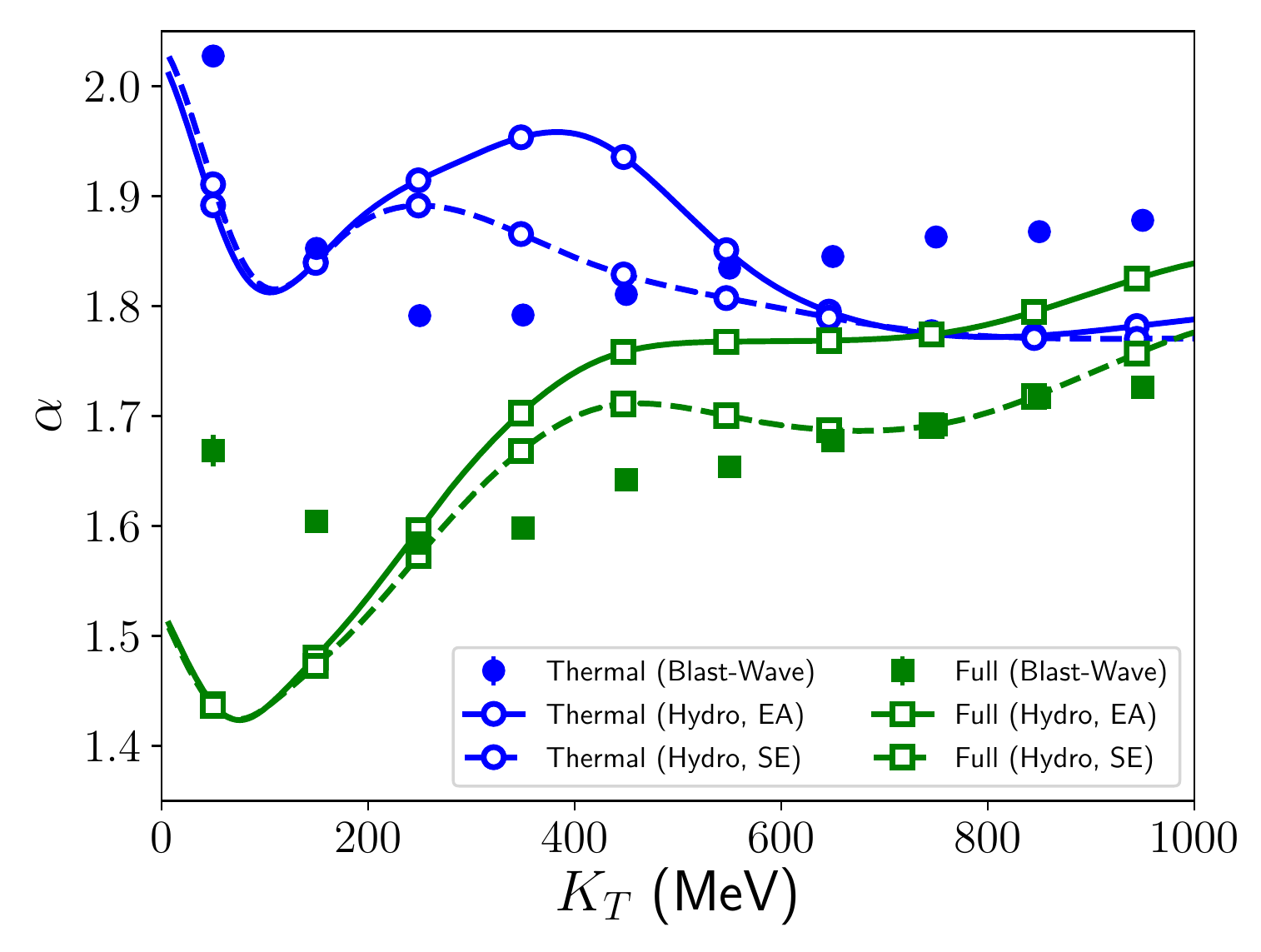}
\caption{The L\'evy parameter of the 3D fit to the correlation function according to Eq.~\eqref{correl_func_Levy_3Ddiag_form}. Compared are simulations with and without resonances.
\label{f:fit3}}
\end{figure}   
%
This is plotted in Fig.~\ref{f:fit3} as a function of $K_T$ for both models. We can see that the obtained $\alpha$'s are closer to 2 than in the case of fitting the one-dimensional correlation functions in $Q_{\mathrm{LCMS}}$, although considerable deviations from 2 are still present. Inclusion of resonance decays lowers $\alpha$ by about 0.1--0.3, depending on $K_T$.

\section{Conclusions}
\label{Conclusions}

In this paper we have presented an analysis of the L\'evy index and how it may be influenced by a variety of different mechanisms which are relevant when measuring the correlation function.  
The deviations appear to be more pronounced in the blast-wave model than in the hydrodynamic model, but the large deviations from $\alpha=2$ nevertheless occur in both approaches explored here.  We thus expect our primary conclusions to be robust and relatively model-independent.

We have found that the most significant deviations arise from two separate sources.  The first source results when projecting the correlation function from its dependence on the full three-dimensional relative momentum $\vec{q}$ to a one-dimensional dependence on the scalar relative momentum $Q$.  We have noted two separate definitions of $Q$ which occur in the literature, and have found that the correlation functions projected against each of these quantities are quite similar to one another.  The effect of projection alone generically leads to values of $\alpha$ in the range 1.3-1.6 for a wide range of $K_T$ values.
This is true even when the three-dimensional correlation function is well described by $\alpha=2$: at low $K_T$ the 3D fit to correlation function only due to thermal pions in Fig.~\ref{f:fit3} gives $\alpha$ near 2, while the projection of the same correlation function in Fig.~\ref{f:res} suppresses $\alpha$ considerably.
Scalar versions of the relative momentum $Q$ therefore tend to obscure some of the underlying physics in a way which makes detailed properties of the full three-dimensional correlation function significantly more difficult to reconstruct.

The second significant source of non-Gaussianity, as represented by the extracted values of the L\'evy index $\alpha$, arises from resonance decays.  In practice, we observe that the extracted value of $\alpha$ is reduced by an additional 0.1-0.2 in models which include resonance decays, when compared with models in which only direct pion production is considered.

In addition to the effects of one-dimensional projection and resonance decays, we also considered the effects of ensemble averaging and bin-averaging in the pair momentum $\vec{K}$.  However, we found these effects to have a relatively small influence on the value of $\alpha$, as compared with the effects of resonance decays and one-dimensional projection.

We have also demonstrated the well known fact that the correlation function tends to be non-Gaussian in the longitudinal direction, just because the source is non-Gaussian due to its boost-invariant expansion. 

Interestingly, we observed that the L\'evy index moved somewhat closer to the Gaussian limit $\alpha=2$ when the full three-dimensional correlation function was parameterized using Eq.~\eqref{correl_func_Levy_3Ddiag_form}.  In this case, the effects of one-dimensional projection are eliminated.  The inclusion of resonance decays still has a visible effect in both models, but the precise $K_T$-dependence of the L\'evy $\alpha$ is now much more sensitive to underlying differences between the two models.  In addition, a comparison of the single-event (SE) and ensemble-averaged (EA) curves suggests that at least some of the non-Gaussianity stems from the effects of ensemble averaging.  Any remaining discrepancies from $\alpha=2$ then result either from the omission from Eq.~\eqref{correl_func_Levy_3Ddiag_form} of the usual off-diagonal terms present in Eqs.~\eqref{correl_func_Gaussian_form} and \eqref{correl_func_Levy_form}, or from the fact that the underlying source itself is not perfectly described by a Gaussian \cite{Wiedemann:1995au}.

The analysis presented here could conceivably be improved in a variety of ways.  For instance, the hydrodynamic approach was used to model a boost-invariant system with specific bulk viscosity $\zeta/s=0$ and no hadronic afterburner phase following chemical freeze-out.  A more sophisticated approach would favor a full, 3+1D hydrodynamic simulation with a realistic parametrization of $(\eta/s)(T)$ and $(\zeta/s)(T)$, and a more realistic treatment of the system following hadronization.  Doing so might, for instance, lead to more significant differences between the correlation functions of a single, fluctuating event and the ensemble average of many events.  Similar comments could be made for the blast-wave model.

Furthermore, one could try to reproduce and understand the observed $K_T$ dependence of the Levy index \cite{Adare:2017vig}, which we did not attempt here. 

However, none of these potential improvements is expected to alter the main conclusion of this work, which is that a high degree of non-Gaussianity arising from a L\'evy parametrization of the correlation function can be understood in terms of various aspects of realistic, femtoscopic analyses.  We find that the relatively mundane sources of non-Gaussianity considered here are more than sufficient to account for the sub-Gaussian L\'evy exponents often extracted in experimental analyses.

The results of our analysis thus strongly suggest that much of the non-Gaussian deviation observed in \cite{Adare:2017vig} arises from a variety of relatively mundane, non-critical sources of generically non-Gaussian behavior.  These sources risk contaminating any signal of genuine critical fluctuations or other intriguing physical phenomena which might be present.  A firm, quantitative handle on the effects of these various sources will therefore be essential before any definitive physical insights can be drawn from measurements of the L\'evy index in heavy-ion collisions.


\acknowledgments


C.P. gratefully acknowledges enlightening conversations with Ulrich Heinz.  The work of C.P. was supported by the CLASH project (KAW 2017-0036).
J.C. and B.T. gratefully acknowledge the support from Czech Science Foundation under Grant No. 17-04505S. B.T. acknowledges the grant VEGA 1/0348/18 (Slovakia). The collaboration was supported by the COST Action CA15213 THOR. We used computing resources from both the Minnesota Supercomputing Institute (MSI) at the University of Minnesota and the Ohio Supercomputer Center \cite{OSC}.


\begin{thebibliography}{999}


\bibitem{Heinz:1999rw} 
			U.~Heinz and B.~V.~Jacak,
			Ann.\ Rev.\ Nucl.\ Part.\ Sci.\  {\bf 49}, 529 (1999).


\bibitem{Lisa:2005dd} 
			M.~A.~Lisa, S.~Pratt, R.~Soltz and U.~Wiedemann,
			Ann.\ Rev.\ Nucl.\ Part.\ Sci.\  {\bf 55}, 357 (2005).

\bibitem{Heinz:1996bs} 
			U.~Heinz,
			in {\it Correlations and Clustering Phenomena in Subatomic Physics}, 
			M.\,N.\,Harakeh, J.\,H.\,Koch, and O.\,Scholten (Eds.), 
			NATO ASI Series B: Physics, Vol. 359 (1997) 137-177 
			[arXiv:nucl-th/9609029].

\bibitem{Adare:2017vig} 
  A.~Adare {\it et al.} [PHENIX Collaboration],
  Phys.\ Rev.\ C {\bf 97}, no. 6, 064911 (2018)
  doi:10.1103/PhysRevC.97.064911
  [arXiv:1709.05649 [nucl-ex]].


\bibitem{Csorgo:2009gb} 
  T.~Cs\"or\H{o},
  PoS \textbf{HIGH-PTLHC08} 027 (2008)
  doi:10.22323/1.076.0027
  [arXiv:0903.0669 [nucl-th]].



\bibitem{HanburyBrown:1954amm} 
			R.~Hanbury Brown and R.~Q.~Twiss,
			Phil.\ Mag.\  {\bf 45}, 663 (1954).
			
\bibitem{Brown:1956zza} 
			R.~H.~Brown and R.~Q.~Twiss,
			Nature {\bf 177}, 27 (1956).

\bibitem{HanburyBrown:1956bqd} 
			R.~Hanbury Brown and R.~Q.~Twiss,
			Nature {\bf 178}, 1046 (1956).

\bibitem{Wiedemann:1999qn} 
			U.~A.~Wiedemann and U.~Heinz,
			Phys.\ Rept.\  {\bf 319}, 145 (1999).

\bibitem{Lisa:2008gf} 
			M.~A.~Lisa and S.~Pratt,
			in {\it Relativistic Heavy-Ion Physics}, R. Stock (ed.), Landolt-B\"ornstein {\bf I 23} (2010),
			Sect. 8.2 [arXiv:0811.1352 [nucl-ex]].

\bibitem{Heinz:2004qz} 
  U.~W.~Heinz,
  hep-ph/0407360.

\bibitem{Pratt:1997pw} 
			S.~Pratt,
			Phys.\ Rev.\ C {\bf 56}, 1095 (1997).


\bibitem{Wiedemann:1998ng} 
  U.~A.~Wiedemann, D.~Ferenc and U.~W.~Heinz,
  Phys.\ Lett.\ B {\bf 449}, 347 (1999)
  doi:10.1016/S0370-2693(99)00087-8
  [nucl-th/9811103].
  

\bibitem{Akkelin:1995gh} 
  S.~V.~Akkelin and Y.~M.~Sinyukov,
  Phys.\ Lett.\ B {\bf 356}, 525 (1995).
  doi:10.1016/0370-2693(95)00765-D



\bibitem{Khachatryan:2010un} 
  V.~Khachatryan {\it et al.} [CMS Collaboration],
  Phys.\ Rev.\ Lett.\  {\bf 105}, 032001 (2010)
  doi:10.1103/PhysRevLett.105.032001
  [arXiv:1005.3294 [hep-ex]].


\bibitem{Aamodt:2010jj} 
  K.~Aamodt {\it et al.} [ALICE Collaboration],
  Phys.\ Rev.\ D {\bf 82}, 052001 (2010)
  doi:10.1103/PhysRevD.82.052001
  [arXiv:1007.0516 [hep-ex]].

\bibitem{Aamodt:2011kd} 
  K.~Aamodt {\it et al.} [ALICE Collaboration],
  Phys.\ Rev.\ D {\bf 84}, 112004 (2011)
  doi:10.1103/PhysRevD.84.112004
  [arXiv:1101.3665 [hep-ex]].

\bibitem{Abelev:2014pja} 
  B.~B.~Abelev {\it et al.} [ALICE Collaboration],
  Phys.\ Lett.\ B {\bf 739}, 139 (2014)
  doi:10.1016/j.physletb.2014.10.034
  [arXiv:1404.1194 [nucl-ex]].


\bibitem{Eggers:2005qz} 
  H.~C.~Eggers, B.~Buschbeck and F.~J.~October,
  AIP Conf.\ Proc.\  {\bf 828}, no. 1, 559 (2006)
  doi:10.1063/1.2197470
  [hep-ex/0511050].

\bibitem{Csorgo:1999wx} 
  T.~Cs\"or\H{o}, A.~T.~Szerzo and S.~Hegyi,
  Phys.\ Lett.\ B {\bf 489}, 15 (2000)
  doi:10.1016/S0370-2693(00)00935-7
  [hep-ph/9912220].

  




\bibitem{Csorgo:2003uv} 
  T.~Cs\"or\H{o}, S.~Hegyi and W.~A.~Zajc,
  Eur.\ Phys.\ J.\ C {\bf 36}, 67 (2004)
  doi:10.1140/epjc/s2004-01870-9
  [nucl-th/0310042].
  
  
\bibitem{Plumberg:2013nga} 
  			C.~J.~Plumberg, C.~Shen and U.~Heinz,
			Phys.\ Rev.\ C {\bf 88}, 044914 (2013)
  			[Erratum: Phys.\ Rev.\ C {\bf 88}, 069901 (2013)].
  			

\bibitem{Plumberg:2015mxa} 
			C.~Plumberg and U.~Heinz,
			Phys.\ Rev.\ C {\bf 92},  044906 (2015)
			[Addendum: Phys.\ Rev.\ C {\bf 92}, 049901 (2015)].

\bibitem{Miskowiec:1997ay} 
  D.~Miskowiec and S.~Voloshin,
  Acta Phys.\ Hung.\ A {\bf 9}, 283 (1999)
  [nucl-ex/9704006].

\bibitem{Wiedemann:1996ig} 
			U.~A.~Wiedemann and U.~Heinz,
			Phys.\ Rev.\ C {\bf 56}, 3265 (1997).


\bibitem{Frodermann:2006sp} 
  E.~Frodermann, U.~Heinz and M.~A.~Lisa,
  Phys.\ Rev.\ C {\bf 73}, 044908 (2006)
  doi:10.1103/PhysRevC.73.044908
  [nucl-th/0602023].



\bibitem{Tomasik:2008fq}
  B. Tom\'a\v{s}ik, 
  Comput.\ Phys.\ Commun.\ {\bf 180} (2009) 1642.
  
  \bibitem{Tomasik:2016skq}
  B. Tom\'a\v{s}ik,
  Comput.\ Phys.\ Commun. {\bf 207} (2016)  545.

\bibitem{Siemens:1978pb}
  P.J.~Siemens,  and J.O.~Rasmussen, 
  Phys.\ Rev.\ Lett. \textbf{42} (1979)   880.

\bibitem{Schnedermann:1993ws}
  E.~Schnedermann, J.~Sollfrank,  and U.~Heinz,
  Phys.\ Rev.\ C \textbf{48} \textbf{1993} 2462.

\bibitem{Csorgo:1995bi}
  T.~Cs\"org\H{o} and B.~L\"orstad,
  Phys.\ Rev.\ C \textbf{54} (1996) 1390.

\bibitem{Tomasik:1999cq}
  B.~Tom\'a\v{s}ik, U.A.~Wiedemann, and U.~Heinz,
  Acta Phys.\ Hung.\ A \textbf{17} (2003) 105.

\bibitem{Retiere:2003kf}
  F.~Reti\'ere  and M.A.~Lisa, 
  Phys.\ Rev.\ C \textbf{70} (2004) 044907.
  
\bibitem{Cooper:1974mv} 
			F.~Cooper and G.~Frye,
			Phys.\ Rev.\ D {\bf 10}, 186 (1974).


\bibitem{Cimerman:2017lmm}
  J. Cimerman,  B.~Tom\'a\v{s}ik, M.~Csan\'ad,  and S.~L\"ok\"os,
  Eur.\ Phys.\ J.\ A \textbf{53} (2017)  161.

\bibitem{CRAB}
S. Pratt: CoRrelation AfterBurner (CRAB), v3, available from https://karman.physics.purdue.edu/oscar/repo/

\bibitem{Song:2007ux} 
  H.~Song and U.~W.~Heinz,
  Phys.\ Rev.\ C {\bf 77}, 064901 (2008)
  doi:10.1103/PhysRevC.77.064901
  [arXiv:0712.3715 [nucl-th]].





\bibitem{Shen:2014vra} 
			C.~Shen, Z.~Qiu, H.~Song, J.~Bernhard, S.~Bass and U.~Heinz,
			Comput.\ Phys.\ Commun.\  {\bf 199}, 61 (2016).

\bibitem{Plumberg:2016sig} 
  C.~Plumberg and U.~Heinz,
  Phys.\ Rev.\ C {\bf 98}, no. 3, 034910 (2018)
  doi:10.1103/PhysRevC.98.034910
  [arXiv:1611.03161 [nucl-th]].

\bibitem{Broniowski:2007nz} 
  W.~Broniowski, M.~Rybczynski and P.~Bozek,
  Comput.\ Phys.\ Commun.\  {\bf 180}, 69 (2009)
  doi:10.1016/j.cpc.2008.07.016
  [arXiv:0710.5731 [nucl-th]].


\bibitem{Alver:2008aq} 
  B.~Alver, M.~Baker, C.~Loizides and P.~Steinberg,
  arXiv:0805.4411 [nucl-ex].


\bibitem{Loizides:2014vua} 
  C.~Loizides, J.~Nagle and P.~Steinberg,
  SoftwareX {\bf 1-2}, 13 (2015)
  doi:10.1016/j.softx.2015.05.001
  [arXiv:1408.2549 [nucl-ex]].

\bibitem{Israel:1979wp} 
  W.~Israel and J.~M.~Stewart,
  Annals Phys.\  {\bf 118}, 341 (1979).
  doi:10.1016/0003-4916(79)90130-1



  
\bibitem{iEBEVISHNUdownload}
   URL: https://u.osu.edu/vishnu/
   
   
\bibitem{Beringer:1900zz} 
  J.~Beringer {\it et al.} [Particle Data Group],
  Phys.\ Rev.\ D {\bf 86}, 010001 (2012).
  doi:10.1103/PhysRevD.86.010001
   
\bibitem{myPDGfile}
			URL: https://raw.githubusercontent.com/~
			astrophysicist87/~
			iEBE-Plumberg/master/EBE-Node/~
			HoTCoffeeh/EOS/pdg.dat

\bibitem{Qiu:2012tm} 
  Z.~Qiu, C.~Shen and U.~W.~Heinz,
  Phys.\ Rev.\ C {\bf 86}, 064906 (2012)
  doi:10.1103/PhysRevC.86.064906
  [arXiv:1210.7010 [nucl-th]].
  
  
  
\bibitem{Wiedemann:1995au} 
  U.~A.~Wiedemann, P.~Scotto and U.~W.~Heinz,
  Phys.\ Rev.\ C {\bf 53}, 918 (1996)
  doi:10.1103/PhysRevC.53.918
  [nucl-th/9508040].


  \bibitem{OSC}
   Ohio Supercomputer Center. 1987. Ohio Supercomputer Center. Columbus OH: Ohio Supercomputer Center. \url{http://osc.edu/ark:/19495/f5s1ph73}.
  


  



\end{thebibliography}
\end{document}